\documentclass[
 aps,
 prd,
 reprint,
 superscriptaddress,
 longbibliography,
 nofootinbib,
 floatfix
]{revtex4-2}

\usepackage[T1]{fontenc}
\usepackage[utf8]{inputenc}
\usepackage{lmodern}
\usepackage{microtype}

\usepackage{amsmath,amssymb,amsfonts,mathtools,bm}
\usepackage{enumitem}
\usepackage{physics}
\usepackage{booktabs}
\usepackage{graphicx}
\usepackage[dvipsnames]{xcolor}

\usepackage{placeins}

\usepackage{hyperref}

\hypersetup{
  colorlinks=true,
  linkcolor=blue!55!black,
  citecolor=blue!55!black,
  urlcolor=blue!55!black,
  filecolor=blue!55!black,
  breaklinks=true
}

\usepackage[nameinlink,noabbrev]{cleveref}

\usepackage{academicons}
\usepackage{orcidlink}

\newcommand{\mpl}{M_{\rm Pl}}
\newcommand{\Nsm}{\mathcal N_{\rm sm}}

\newcommand{\dtot}{d_{\rm tot}}
\newcommand{\dek}{d_{\rm ek}}
\newcommand{\dconv}{d_{\rm conv}}
\newcommand{\db}{d_{\rm b}}
\newcommand{\dpost}{d_{\rm post}}
\newcommand{\dallowed}{d_{\rm allowed}}

\newcommand{\ABKL}{\mathcal A_{\rm BKL}}
\newcommand{\deff}{d_{\rm eff}}
\newcommand{\kappag}{\kappa_g}
\newcommand{\Thetareq}{\Theta_{\rm req}}

\newcommand{\sigmadot}{\dot\sigma}
\newcommand{\Ordo}{\mathcal O}

\begin{document}

\title{Phase-resolved field-space distance criteria in ekpyrotic, bouncing and cyclic cosmologies}

\author{Marcin Postolak\orcidlink{0000-0003-4868-6358}}
\email[]{marcin.postolak@uwr.edu.pl,postolak.marcin@gmail.com}
\affiliation{Institute of Theoretical Physics, Faculty of Physics and Astronomy, \href{https://ror.org/00yae6e25}{University of Wrocław}, pl. M. Borna 9, 50-204 Wrocław, Poland}


\begin{abstract}
The inflationary Lyth bound relates the primordial tensor amplitude to the inflaton field excursion. In ekpyrotic, bouncing and cyclic cosmologies there is no analogous universal tensor-to-field-distance relation because scalar and tensor perturbations depend on entropy conversion, matching through the bounce, and the specific mechanism that violates or evades the null energy condition. Here we propose a phase-resolved criterion for the accumulated scalar field trajectory length in a non-inflationary smoothing history. The quantity constrained in this study is not, in general, the geodesic distance between the initial and final field-space points. Rather, it is the invariant path length accumulated along the actual cosmological trajectory. Therefore, the resulting inequalities should be understood as sufficient, conservative trajectory length criteria for small-field control, not as model-independent necessary exclusions based on geodesic distance swampland reasoning. We also impose BKL (Belinski-Khalatnikov-Lifshitz) anisotropy suppression as an additional constraint on the ekpyrotic phase. In the canonical phase of the ekpyrotic contraction, we recover the known small-field scaling and embed it in a total trajectory length budget inequality. We impose three requirements: a BKL anisotropy suppression that is parameterized separately, a phenomenological cutoff-corrected distance budget inspired by tower of states logic, and observational conversion windows from residual isocurvature and non-Gaussianity. Furthermore, we propose a new master formula that provides a conditional lower limit on the value of the parameter $\epsilon_{\rm ek}$ that depends on the remaining distance available after conversion and the cosmological bounce. We also derive a curvature constraint for scale-invariant entropy perturbations in curved field space which shows that the small total distance and the observed red tilt seem to indicate ultra-fast-roll ekpyrosis, sharp turns, short or strongly modified bounces, and/or significant negative sectional curvature of the scalar manifold. Finally, we demonstrate methods for testing the distance budget against observational data. Examples include $n_s$ running, primordial non-Gaussianity, isocurvature, a stochastic GWs background (SGWB), and late-time dark energy distance constraints. The observational discussion is organized into explicit diagnostic maps, which are used to identify and analyze specific features and patterns in the data. We also apply the formalism to exponential and cyclic ekpyrotic scalar field potentials, providing a mechanism-resolved interpretation of the bounce or crossover contribution for several specific realizations.
\end{abstract}

\maketitle

\section{Introduction}

The Lyth bound \cite{Lyth:1996im} is one of the cleanest kinematic connections between an observable quantity and a microscopic field-space distance in early Universe cosmology. In canonical single-field slow-roll inflation, the relations:
\begin{equation}\label{r-epsilon}
    r=16\epsilon \quad\land\quad \frac{d\phi}{dN}=\mpl\sqrt{2\epsilon}
\end{equation}
imply:
\begin{equation}
 \frac{\Delta\phi}{\mpl}=\int dN\sqrt{\frac{r(N)}{8}}\,,
 \label{eq:lyth-standard}
\end{equation}
so that a detectable primordial tensor amplitude generically points to a Planckian or super-Planckian inflaton excursion \cite{Lyth:1996im}. The relation has been generalized within the effective theory of single-field inflation, where nontrivial dynamics does not automatically eliminate field-range constraints \cite{Baumann:2011ws}.

Non-inflationary alternatives such as ekpyrotic, bouncing and cyclic cosmologies address the flatness, horizon and anisotropy problems by a phase of slowly contracting ultra-stiff evolution rather than accelerated expansion \cite{Khoury:2001wf,Steinhardt:2001st,Lehners:2008vx,Battefeld:2014uga,Postolak:2024xtm}. In such scenarios, the role of tensor perturbations is fundamentally different from canonical slow-roll inflation. In minimal ekpyrotic models, the tensor spectrum is typically blue and negligible on cosmic microwave background (CMB) scales \cite{Boyle:2003km,Lehners:2008vx}, whereas special constructions, such as S-brane models, may produce observable tensors with model-dependent consistency relations \cite{Brandenberger:2020tcr}. In non-singular modified gravity bounces, scalar and tensor spectra depend on the detailed bounce sector and transfer functions \cite{Ilyas:2020zcb,Zhu:2021whu}. Consequently, there is no model-independent ekpyrotic identity analogous to \eqref{r-epsilon}.

However, the absence of a universal tensor-to-field range relation does not mean that field distance is irrelevant. Ekpyrotic smoothing itself has a simple field-range cost. Lehners demonstrated that imposing sub-Planckian field excursions can drive scale-free ekpyrosis to the ultra-fast-roll regime, which can lead to observationally relevant running \cite{Lehners:2018vgi}. This study takes a complementary step. We propose that the natural, non-inflationary analogue of Lyth's logic is a \textit{phase-resolved field-space distance budget} for the entire cosmological history:
\begin{equation}
 \dtot=\dek+\dconv+\db+\dpost\,.
 \label{eq:total-budget-intro}
\end{equation}
In this expression, \(d_{\rm tot}\) denotes the accumulated invariant
trajectory length along the physical cosmological path in field space. It
should not be confused with the geodesic distance between the initial and
final field-space points. Since the endpoint geodesic distance is always less than or equal to the path length, a trajectory length criterion is stronger than a geodesic-distance criterion. Consequently, failure to satisfy the trajectory length budget does not by itself exclude a model unless the specific EFT cutoff or UV completion is known to depend on the accumulated path length.

The point is simple. If an effective field theory (EFT) is required to remain within a finite field-space distance, then ekpyrotic smoothing, entropy conversion, the bounce, and post-bounce matching all compete for the same distance.

\subsection*{Relation to previous studies}

The ekpyrotic smoothing contribution in equation \eqref{eq:dek-general} is not claimed to be new. It is closely related to the small-field ekpyrotic argument in \cite{Lehners:2018vgi}, in which imposing a sub-Planckian field range was shown to drive scale-free ekpyrosis toward ultra-fast-roll. The new contribution of this work is extending this one-phase bound to a complete non-inflationary history, including BKL anisotropy suppression, entropy-to-curvature conversion, bounce duration, cutoff loss, and post-bounce evolution. Thus, the result is not an ekpyrotic tensor-to-scalar Lyth bound but rather a field-space distance-budget criterion.

The main results are as follows: First, we rederive the ekpyrotic smoothing contribution in terms of the growth of the comoving Hubble scale. Second, we introduce conditional lower limits for conversion and bounce. Third, we combine these criteria into a master inequality and solve for the minimum ekpyrotic fast-roll parameter, $\epsilon_{\rm ek}$. Fourth, we connect the same distance budget to observable quantities, particularly $n_s$, non-Gaussianity, residual isocurvature, CMB B modes, and late-time scalar field evolution, as suggested by recent dark energy data.

\begin{figure*}[htbp]
 \centering
 \includegraphics[width=1\textwidth]{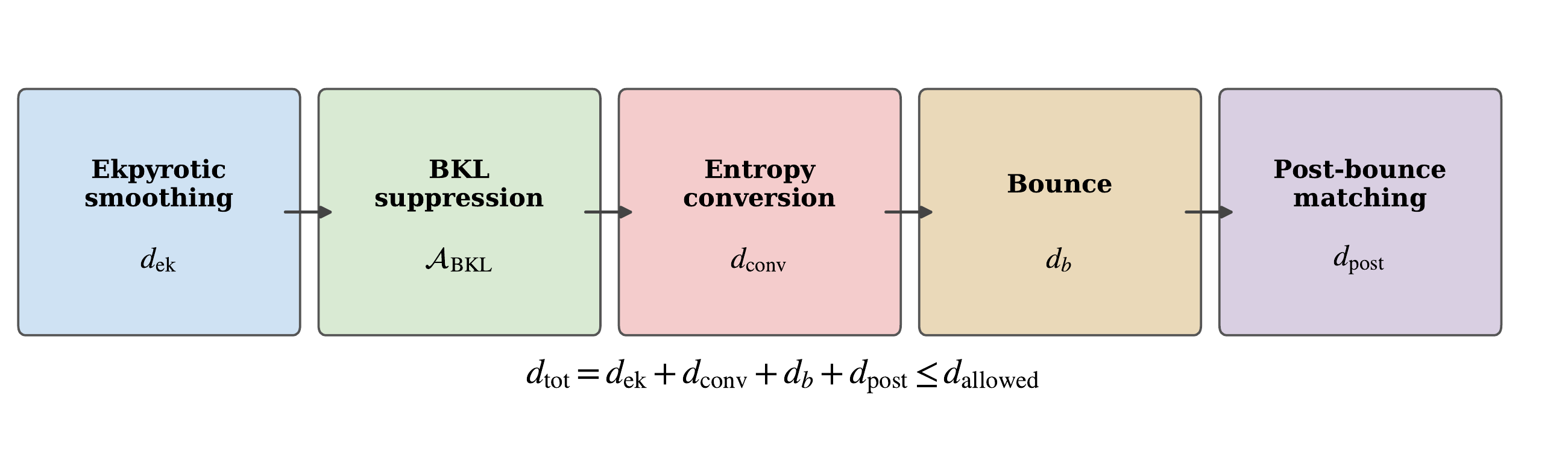}
 \caption{Phase-resolved field-space distance budget for a complete non-inflationary smoothing history. Ekpyrotic smoothing, anisotropy suppression, entropy conversion, bounce dynamics and post-bounce evolution contribute to the accumulated field-space path length.}
 \label{fig:distance-budget-flow}
\end{figure*}

\section{Invariant field-space distance}
\label{sec:distance}

Let us consider $n$ scalar fields that are minimally coupled to Einstein's theory of general relativity (GR):
\begin{equation}
\begin{aligned}
    S= & \int d^4x\sqrt{-g}\left[\frac{\mpl^2}{2}R-\frac{1}{2}G_{IJ}(\phi)g^{\mu\nu}\partial_\mu\phi^{I}\partial_\nu\phi^{J}-V(\phi)\right] \\
    & +S_{\rm{b/EFT}}\,,
\end{aligned}
\label{eq:action}
\end{equation}
where $G_{IJ}$ is the field-space metric, and $S_{\rm b/EFT}$ denotes NEC-violating operators, as well as higher-derivative or modified gravity operators, which are relevant near the bounce. Using an invariant field-space metric and covariant field-space distance follows the standard multifield formalism \cite{Gordon:2000hv,Gong:2011uw,Renaux-Petel:2015mga}. 

The following assumptions are used throughout:
\begin{enumerate}[label=(\roman*)]
    \item Einstein-scalar dynamics are assumed during the ekpyrotic smoothing phase but not necessarily during the bounce;
    \item The bounce sector is treated kinematically through the effective potential, the bounce action, and the bounce interval, rather than being derived from a specific microscopic Lagrangian;
    \item Entropy perturbations are treated in the approximate decoupling limit during their generation;
    \item Observational conversion criteria are interpreted as phenomenological consistency windows rather than as model-independent numerical limits.
\end{enumerate}

In the case of spatially flat FLRW background, the line element becomes of the form:
\begin{equation}
 ds^2=-dt^2+a^2(t)\,d\vec{x}^2\,,
\end{equation}
so that one can define the field-space speed:
\begin{equation}
 \sigmadot^2\equiv G_{IJ}\dot\phi^{I}\dot\phi^{J}\,.
\end{equation}
The invariant scalar distance is:
\begin{equation}
 \Delta s=\int ds_{\rm{field}}=\int\sigmadot\,dt\,,
\end{equation}
and the dimensionless distance takes the form of:
\begin{equation}
 d\equiv\frac{\Delta s}{\mpl}=\frac{1}{\mpl}\int\sigmadot\,dt\,.
 \label{eq:d-def}
\end{equation}
It is important to distinguish the \textbf{\textit{trajectory length}} used in this study from the \textit{geodesic distance between the endpoints of the trajectory}. The quantity:
\begin{equation}
    d_{\rm traj}\equiv\frac{1}{\mpl}\int_{\gamma} ds_{\rm field}=\frac{1}{\mpl}\int\dot\sigma\,dt
\end{equation}
is the \textbf{length of the actual cosmological path} $\boldsymbol{\gamma}$ \textbf{in the field space}. On the other hand, the geodesic distance between the initial and final field-space points takes the form of:
\begin{equation}
    d_{\rm geo}\bigl(\phi_i,\phi_f\bigr)\equiv\frac{1}{\mpl}\inf_{\Gamma:\,\phi_{i}\to\phi_{f}}\int_{\Gamma} ds_{\rm field}\,,
\end{equation}
where $\inf_{\Gamma:\,\phi_{i}\to\phi_{f}}$ denotes the infimum over all
field-space curves $\Gamma$ connecting the initial point $\phi_{i}$
to the final point $\phi_{f}$. Thus, $d_{\rm geo}$ is the shortest
possible field-space distance between the endpoints and:
\begin{equation}
    d_{\rm traj}=\frac{1}{\mpl}\int_\gamma ds_{\rm field}
\end{equation}
is the length of the actual cosmological trajectory $\gamma$. Therefore:
\begin{equation}
 d_{\rm geo}\bigl(\phi_i,\phi_f\bigr)\leq d_{\rm traj}\,.
\end{equation}
Equality only occurs when the cosmological trajectory is a geodesic segment, up to reparametrization. Specifically, a winding or strongly curved trajectory may have a significant accumulated path length, while its endpoint geodesic distance remains small.

Away from the bounce, assuming ordinary Einstein-scalar dynamics, the Friedmann equations yield:
\begin{equation}
 \dot H=-\frac{\sigmadot^2}{2\mpl^2}
 \quad\land\quad
 \epsilon\equiv -\frac{\dot H}{H^2}=\frac{\sigmadot^2}{2\mpl^{2}H^2}\,.
 \label{eq:eps-def}
\end{equation}
Thus:
\begin{equation}
 \sigmadot=\mpl \left|H\right|\sqrt{2\epsilon}\,,
\end{equation}
and\footnote{In the remainder of the paper, unless stated otherwise, $d$ denotes the accumulated trajectory length $d_{\rm traj}$, not the endpoint geodesic distance $d_{\rm geo}$.}:
\begin{equation}
 d=\int \left|H\right|dt\sqrt{2\epsilon}\,.
 \label{eq:d-Hdt}
\end{equation}
This is the basic kinematic identity underlying the non-inflationary distance budget.

\section{Ekpyrotic smoothing distance}
\label{sec:ekpyrotic-distance}

We know that during the ekpyrotic contraction:
\begin{equation}
 H<0
 \quad\land\quad
 \epsilon>3
 \quad\land\quad
 \omega=\frac{2}{3}\epsilon-1\gg1\,.
\end{equation}
The relevant smoothing measure is not the expansion e-fold number:
\begin{equation}
    N=\ln{a}\,,
\end{equation}
but the growth of the comoving Hubble scale:
\begin{equation}
 \Nsm\equiv\ln{\bigl(a\left|H\right|\bigr)}\,.
\end{equation}
Its differential is:
\begin{align}
 d\Nsm
 &=d\ln{a}+d\ln{\left|H\right|} \\
 &=H\,dt+\frac{\dot{H}}{H}dt \\
 &=\left(1-\epsilon\right)H\,dt\,.
\end{align}
Since $H<0$ and $\epsilon>1$, one can conclude that:
\begin{equation}
 d\Nsm=\left(\epsilon-1\right)\left|H\right|\,dt\,.
 \label{eq:dNsm}
\end{equation}
Using \eqref{eq:d-Hdt}, the ekpyrotic distance is therefore:
\begin{equation}
 \dek=\int d\Nsm\,\frac{\sqrt{2\epsilon}}{\left(\epsilon-1\right)}\,,
 \label{eq:dek-general}
\end{equation}
which is the ekpyrotic analogue of the kinematic part of the Lyth bound. For constant $\epsilon_{\rm ek}$:
\begin{equation}
 \dek\simeq\frac{\sqrt{2\epsilon_{\rm ek}}}{\left(\epsilon_{\rm ek}-1\right)}\,\Nsm\,.
 \label{eq:dek-constant}
\end{equation}
In the ultra-ekpyrotic limit $\epsilon_{\rm ek}\gg1$:
\begin{equation}
 \dek\simeq\sqrt{\frac{2}{\epsilon_{\rm ek}}}\,\Nsm\,.
 \label{eq:dek-ultra}
\end{equation}
Hence, the ekpyrotic smoothing phase alone is sub-Planckian only if:
\begin{equation}
 \epsilon_{\rm ek}\gtrsim 2\Nsm^2\,.
 \label{eq:eps-basic}
\end{equation}
For $\Nsm\simeq 60$, this gives:
\begin{equation}
 \epsilon_{\rm ek}\gtrsim 7200
 \quad\land\quad
 \omega_{\rm ek}\gtrsim 4800\,.
\end{equation}
This reproduces the basic ultra-fast-roll implication of small-field ekpyrosis \cite{Lehners:2018vgi}. The novelty developed below is to make the criterion phase-resolved.

\section{Anisotropy suppression and a BKL distance criterion}
\label{sec:bkl}

Ekpyrotic contraction is necessary for more than just shrinking the comoving Hubble radius, namely, it must also suppress the shear anisotropy underlying the Belinski-Khalatnikov-Lifshitz instability near spacelike singularities \cite{Belinsky:1970ew,Erickson:2003zm}. This yields a second, closely related distance criterion.

For an approximately homogeneous but anisotropic background, the shear contribution scales as:
\begin{equation}
 \rho_{\sigma}\propto a^{-6}\,.
\end{equation}
For an ekpyrotic scalar with a nearly constant equation of state:
\begin{equation}
 \rho_{\phi}\propto a^{-2\epsilon}\,,
\end{equation}
so that:
\begin{equation}
 \frac{\rho_{\sigma}}{\rho_\phi}\propto a^{2\epsilon-6}\,.
\end{equation}
Since $a$ decreases during contraction, this ratio is suppressed only for:
\begin{equation}
    \epsilon>3\,,
\end{equation}
which is the usual ultra-stiff ekpyrotic condition. This is the same physical mechanism by which ekpyrotic contraction suppresses mixmaster-type chaotic behavior in the:
\begin{equation}
    \omega>1
\end{equation}
regime \cite{Erickson:2003zm}.

Now, let us define the logarithmic anisotropy suppression requirement of the form:
\begin{equation}
 \ABKL\equiv\ln{\left[\frac{\left(\frac{\rho_{\sigma}}{\rho_{\phi}}\right)_{i}}{\left(\frac{\rho_{\sigma}}{\rho_{\phi}}\right)_{f}}\right]}\,.
 \label{eq:ABKL-def}
\end{equation}
For a constant $\epsilon$ value one gets:
\begin{equation}
 \ABKL=2\left(\epsilon-3\right)\,\left|\Delta N\right|\,,
 \label{eq:ABKL-DN}
\end{equation}
where:
\begin{equation}
    \left|\Delta N\right|=\ln{\left(\frac{a_{i}}{a_{f}}\right)}\,.
\end{equation}
The same phase is described by the exact relations:
\begin{equation}
 \Nsm=\left(\epsilon-1\right)\,\left|\Delta N\right|
 \quad\land\quad
 \dek=\sqrt{2\epsilon}\,\left|\Delta N\right|\,.
\end{equation}
Eliminating $\left|\Delta N\right|$ using \eqref{eq:ABKL-DN}, the distance needed to achieve a prescribed BKL suppression is:
\begin{equation}
 \dek=\frac{\ABKL\sqrt{2\epsilon}}{2\left(\epsilon-3\right)}\,.
 \label{eq:d-BKL}
\end{equation}
For $\epsilon\gg1$ it yields:
\begin{equation}
 \dek\simeq\frac{\ABKL}{\sqrt{2\epsilon}}\,.
 \label{eq:d-BKL-ultra}
\end{equation}
If only the distance $d_{\rm rem}$ is available for the ekpyrotic smoothing phase, then the exact BKL condition is given by:
\begin{equation}
 \frac{\ABKL\sqrt{2\epsilon}}{2(\epsilon-3)}\le d_{\rm rem}\,.
\end{equation}
Introducing the quantity:
\begin{equation}
    x=\sqrt{\epsilon}
\end{equation}
gives:
\begin{equation}
 d_{\rm rem}\,x^2-\frac{\ABKL}{\sqrt{2}}x-3d_{\rm rem}\ge 0\,,
\end{equation}
and therefore:
\begin{equation}
 \epsilon_{\rm ek}\ge\left[\frac{\ABKL}{2\sqrt2\,d_{\rm rem}}
 +\sqrt{3+\frac{\ABKL^2}{8d_{\rm rem}^2}}\right]^2\,.
 \label{eq:eps-BKL-exact}
\end{equation}
In the strong suppression limit:
\begin{equation}
    \ABKL\gg d_{\rm rem}
\end{equation}
one obtains the compact result of:
\begin{equation}
 \epsilon_{\rm ek}\gtrsim\frac{\ABKL^2}{2d_{\rm rem}^2}\,.
 \label{eq:eps-BKL-approx}
\end{equation}
Combining the smoothing and anisotropy requirements gives the conservative criterion:
\begin{equation}
 \epsilon_{\rm ek}\gtrsim\max\left[\frac{2\Nsm^2}{d_{\rm rem}^2},
 \frac{\ABKL^2}{2d_{\rm rem}^2}\right]\,.
 \label{eq:eps-smoothing-BKL-max}
\end{equation}
In the simplest constant $\epsilon$ model, the two terms are not independent because, at large values of $\epsilon$:
\begin{equation}
    \ABKL=\frac{2\left(\epsilon-3\right)\Nsm}{\left(\epsilon-1\right)}\simeq 2\Nsm\,.
\end{equation}
However, they differ when the required anisotropy suppression is determined by initial shear, spatial curvature, or a non-constant ekpyrotic profile. In this sense, the BKL limit converts the distance budget into a criterion that simultaneously smooths and isotropizes.

\section{Total distance budget}
\label{sec:budget}

Let us introduce the notion of the \textit{\textbf{total invariant trajectory length (accumulated field-space path length)}}:
\begin{equation}
 \dtot=\dek+\dconv+\db+\dpost\,,
 \label{eq:budget}
\end{equation}
where $\dconv$ is the entropy-to-curvature conversion distance, $\db$ is the bounce distance, and $\dpost$ denotes post-bounce matching, reheating or relaxation. The phase-resolved budget proposed here is a constraint on the accumulated trajectory length, not directly on the endpoint geodesic distance. A small-field EFT-control criterion is:
\begin{equation}
 \dtot\le d_{\max}\,,
 \label{eq:dmax}
\end{equation}
where $d_{\max}\sim\Ordo(1)$ corresponds to a Planckian field-space radius.
More generally, the maximum usable distance may be smaller than $d_{\rm max}$ if the EFT cutoff decreases along the trajectory. 

The motivation for swampland-distance should be interpreted with care. The usual \textit{\textbf{tower of states}} logic \cite{Ooguri:2006in,Ooguri:2018wrx,Palti:2019pca} is naturally phrased in terms of a geodesic distance in scalar field space, or in terms of the geodesic distance to an infinite-distance locus. It does not, by itself, imply that the EFT cutoff must decrease exponentially with the accumulated length of an arbitrary winding trajectory. Therefore, the path-length version used below should be regarded as a stronger, conservative EFT-control criterion rather than as a necessary consequence of the distance conjecture. A geodesic version of the cutoff condition would be of the following form \cite{Ooguri:2006in,Ooguri:2018wrx,Palti:2019pca}:
\begin{equation}
    \Lambda_{\rm EFT}(\phi)=\Lambda_{0}\exp{\Bigl[-\gamma\,d_{\rm geo}\bigl(\phi_{0},\phi\bigr)\Bigr]}\,,
\end{equation}
and hence:
\begin{equation}
    H_{\rm max}<\min_{t\in\gamma}\Lambda_{\rm EFT}\bigl[\phi(t)\bigr]\,.
\end{equation}
Since:
\begin{equation}
    d_{\rm geo}\bigl[\phi_{i},\phi(t)\bigr]\leq d_{\rm traj}(t)\,,
\end{equation}
imposing an upper limit on the accumulated trajectory length is sufficient to keep all points on the path within the corresponding geodesic ball, but the converse is not true. Thus, violation of the trajectory length budget does not by itself exclude a model unless the specific UV completion or EFT cutoff is shown to depend on accumulated path length.

In what follows, $d_{\rm allowed}$ defined as:
\begin{equation}
    \dallowed\equiv\min\left[d_{\rm max},\frac{1}{\gamma}\ln\left(\frac{\Lambda_{0}}{H_{\rm max}}\right)\right]\,.
\label{eq:dallowed}
\end{equation}
should therefore be understood as an \textbf{assumed trajectory length control scale}. The resulting inequalities provide \textit{\textbf{sufficient limits for a controlled small-trajectory completion, not model-independent necessary conditions}}. This cutoff-corrected budget is particularly important for high-curvature, non-singular bounces. Increasing the bounce scale decreases the available field-space distance, unless the microscopic cutoff is parametrically greater.

Now, we can define:
\begin{equation}
 d_{\rm aux}\equiv \dconv+\db+\dpost
 \quad\land\quad
 d_{\rm rem}\equiv \dallowed-d_{\rm aux}\,.
 \label{eq:drem}
\end{equation}
The ekpyrotic phase is possible only if:
\begin{equation}
    d_{\rm rem}>0
\end{equation}
and:
\begin{equation}
 \frac{\sqrt{2\epsilon_{\rm ek}}}{\left(\epsilon_{\rm ek}-1\right)}\Nsm\le d_{\rm rem}\,.
 \label{eq:eps-ineq-start}
\end{equation}
Assuming that:
\begin{equation}
    x\equiv\sqrt{\epsilon_{\rm ek}}>1
\end{equation}
yields:
\begin{equation}
 \frac{\sqrt2\,\Nsm\,x}{x^{2}-1}\le d_{\rm rem}\,,
\end{equation}
or, equivalently:
\begin{equation}\label{equation1}
 d_{\rm rem}\,x^2-\sqrt2\,\Nsm\,x-d_{\rm rem}\ge0\,.
\end{equation}
The positive root of \eqref{equation1} is given by the following formula:
\begin{equation}
 x\ge\frac{\sqrt2\,\Nsm+\sqrt{2\Nsm^2+4d_{\rm rem}^2}}{2d_{\rm rem}}\,.
\end{equation}
so that:
\begin{equation}
 \epsilon_{\rm ek}\ge\left[\frac{\Nsm}{\sqrt2\,d_{\rm rem}}
 +\sqrt{1+\frac{\Nsm^2}{2d_{\rm rem}^2}}\right]^2\,.
 \label{eq:eps-exact-drem}
\end{equation}
In the limit of:
\begin{equation}
    \Nsm\gg d_{\rm rem}\,,
\end{equation}
we obtain the more compact form:
\begin{equation}
 \epsilon_{\rm ek}\gtrsim\frac{2\Nsm^2}{d_{\rm rem}^2}\,,
 \label{eq:eps-approx-drem}
\end{equation}
with:
\begin{equation}
 d_{\rm rem}=\dallowed-\dconv-\db-\dpost\,.
\end{equation}
This result is the simplest expression of the distance-budget effect. Conversion and bounce costs decrease the available field-space distance for smoothing. This increases the required ekpyrotic fast-roll parameter.

For the exponential ekpyrotic potential:
\begin{equation}
 V(\phi)=-V_0e^{-c\phi/\mpl}\,,
\end{equation}
with the scaling solution:
\begin{equation}
 \epsilon_{\rm ek}=\frac{c^2}{2}\,,
\end{equation}
the inequality \eqref{eq:eps-approx-drem} produces:
\begin{equation}
 c\gtrsim \frac{2\Nsm}{d_{\rm rem}}\,.
 \label{eq:c-bound}
\end{equation}
Thus, a complete sub-Planckian history with $\dallowed=1$, $\Nsm=60$, and $d_{\rm aux}=0.5$ requires:
\begin{equation}
 \epsilon_{\rm ek}\gtrsim 2.88\times10^{4},
 \quad\land\quad
 c\gtrsim 240\,,
\end{equation}
which is \textit{four times stronger} in $\epsilon$ than the ekpyrotic phase only estimate.

\section{Entropy-conversion distance}
\label{sec:conversion}

In many ekpyrotic models, curvature perturbations are generated from the entropy perturbations \cite{Notari:2002yc,Lehners:2007ac,Buchbinder:2007ad}. Therefore, one usually introduce the tangent and normal unit vectors:
\begin{equation}
 T^I=\frac{\dot\phi^I}{\sigmadot}
 \quad\land\quad
 G_{IJ}T^I T^J=1\,,
\end{equation}
and:
\begin{equation}
 G_{IJ}N^I N^J=1
 \quad\land\quad
 G_{IJ}T^I N^J=0\,.
\end{equation}
On the other hand, the covariant turn rate is defined by the relation:
\begin{equation}
 D_{t}T^I=\Omega\,N^I\,.
\end{equation}
The total bending angle is:
\begin{equation}
 \Theta_{\rm conv}=\int \Omega\,dt\,.
\end{equation}
The geodesic curvature of the field-space trajectory is:
\begin{equation}
 \kappag\equiv\frac{\Omega}{\sigmadot}\,.
\end{equation}
Since:
\begin{equation}
    ds_{\rm field}=\sigmadot dt\,,
\end{equation}
then:
\begin{equation}
 d\Theta=\Omega\,dt=\kappag\,ds_{\rm field}\,.
\end{equation}
Therefore:
\begin{equation}
 \Theta_{\rm conv}=\int \kappag\,ds_{\rm field}\,.
\end{equation}
If $\kappag\le \kappa_{g,{\rm max}}$, then:
\begin{equation}
 \Theta_{\rm conv}\le \kappa_{g,{\rm max}}\,\Delta s_{\rm conv}\,.
\end{equation}
Thus, a required bending angle $\Theta_{\rm req}$ imposes the relationship:
\begin{equation}
 \dconv\ge\frac{\Thetareq}{\mpl\,\kappa_{g,{\rm max}}}\,.
 \label{eq:dconv-bound}
\end{equation}
Equivalently, small distance conversion requires:
\begin{equation}
 \mpl\,\kappa_{g,{\rm max}}\gtrsim\frac{\Thetareq}{\dconv}\,.
\end{equation}
The physical interpretation is transparent: efficient conversion in a sub-Planckian interval requires a sharp bend or a small radius of curvature:
\begin{equation}
 R_{\rm turn}=\kappag^{-1}\lesssim \frac{\mpl\,\dconv}{\Thetareq}\,.
\end{equation}

It is also useful to express the same statement in terms of Hubble units. By defining:
\begin{equation}
 \eta_\perp\equiv\frac{\Omega}{\left|H\right|}
 \quad\land\quad
 dn\equiv\left|H\right|\,dt\,.
\end{equation}
one can write:
\begin{equation}\label{theta-conv}
 \Theta_{\rm conv}=\int \eta_\perp\,dn\,.
\end{equation}
For:
\begin{equation}
    \left|\eta_{\perp}\right|\le\eta_{\perp,{\rm max}}
\end{equation}
we obtain:
\begin{equation}
 n_{\rm conv}\ge \frac{\Thetareq}{\eta_{\perp,{\rm max}}}\,.
\end{equation}
On the other hand:
\begin{equation}\label{d-conv}
 \dconv=\int\sqrt{2\epsilon}\,dn\,,
\end{equation}
and for:
\begin{equation}
    \epsilon\ge\epsilon_{{\rm conv},\min}
\end{equation}
we conclude that:
\begin{equation}
 \dconv\ge \sqrt{2\epsilon_{{\rm conv},\min}}\,n_{\rm conv}\,.
\end{equation}
Combining both criteria gives the following constraint:
\begin{equation}
 \Thetareq\le\frac{\eta_{\perp,{\rm max}}}{\sqrt{2\epsilon_{{\rm conv},\min}}}\dconv\,.
 \label{eq:turn-duration-bound}
\end{equation}
Therefore, a small-field scenario cannot simultaneously have a very long, weakly turning and efficient conversion phase.

\subsection{Conversion windows from isocurvature and non-Gaussianity}
\label{sec:conversion-window}

The conversion distance is more than just a theoretical cost, it can also be limited by residual isocurvature and non-Gaussianity. In a two-field transfer formalism one can write:
\begin{equation}
 \mathcal R_f=\mathcal R_i+T_{\mathcal RS}\,\mathcal S_i
 \quad\land\quad
 \mathcal S_f=T_{\mathcal SS}\mathcal S_i\,.
 \label{eq:transfer-matrix}
\end{equation}
Such a transfer matrix approach is a standard formalism in multifield perturbation theory
\cite{Gordon:2000hv,Malik:2008im}. A schematic residual isocurvature fraction is:
\begin{equation}
 \beta_{\rm iso}\simeq\frac{T_{\mathcal SS}^2}{T_{\mathcal RS}^2+T_{\mathcal SS}^2}\,.
 \label{eq:beta-iso}
\end{equation}
In correlated mixed models, Planck bounds the non-adiabatic contribution at the percent level. The precise number depends on the isocurvature mode and the specific correlation choice \cite{Planck:2018jri}. If $\beta_{\rm iso}<\beta_{\rm max}$, then:
\begin{equation}
 \frac{\left|T_{\mathcal RS}\right|}{\left|T_{\mathcal SS}\right|}\gtrsim
 \sqrt{\frac{1-\beta_{\rm max}}{\beta_{\rm max}}}=\sqrt{\frac{1}{\beta_{\rm max}}-1}\,.
 \label{eq:TRS-TSS-bound}
\end{equation}
On superhorizon scales, the curvature perturbation is sourced by the turn rate, schematically:
\begin{equation}
    \frac{d\mathcal R}{dn}\simeq2\eta_\perp\mathcal S\,.
\end{equation}
If $|\eta_\perp|\le\eta_{\perp,{\rm max}}$ and the entropy transfer during conversion is summarized by the residual factor $|T_{\mathcal SS}|$, a conservative conversion efficiency requirement is:
\begin{equation}
 \Delta n_{\rm conv}\gtrsim\frac{\left|T_{\mathcal SS}\right|}{2\eta_{\perp,{\rm max}}}\sqrt{\frac{1}{\beta_{\rm max}}-1}\,.
\end{equation}
Using \eqref{d-conv} gives:
\begin{equation}
 \dconv\gtrsim\frac{\sqrt{2\epsilon_{{\rm conv},\min}}}{2\eta_{\perp,{\rm max}}}\sqrt{\frac{1}{\beta_{\rm max}}-1}\left|T_{\mathcal SS}\right|\,.
 \label{eq:dconv-isocurvature}
\end{equation}
However, the formula \eqref{eq:dconv-isocurvature} should not be read as a universal relation; it is a compact way of saying that CMB adiabaticity imposes a lower cost on conversion once a maximum turn rate is taken into account.

Non-Gaussianity pushes in the opposite direction. A sharp conversion of entropy to curvature can generate significant local-type non-Gaussianity in ekpyrotic and cyclic models \cite{Lehners:2010fy}, though Planck detects no primordial non-Gaussianity signal \cite{Planck:2019kim}.

In a given conversion model, this can be represented as an upper criterion of the form:
\begin{equation}
    \left|\eta_\perp\right|\le\eta_{\perp,{\rm NG}}\,.
\end{equation}
Since $\Theta_{\rm req}$ is given by the relation \eqref{theta-conv}, one obtains:
\begin{equation}
 \Delta n_{\rm conv}\ge\frac{\Theta_{\rm req}}{\eta_{\perp,{\rm NG}}}\,,
\end{equation}
and therefore:
\begin{equation}
 \dconv\gtrsim\sqrt{2\epsilon_{{\rm conv},\min}}\frac{\Theta_{\rm req}}{\eta_{\perp,{\rm NG}}}\,.
 \label{eq:dconv-NG}
\end{equation}
Equations \eqref{eq:dconv-isocurvature} and \eqref{eq:dconv-NG} define an observational conversion window. Isocurvature requires efficient conversion, whereas non-Gaussianity limits the sharpness of the conversion.

\section{Kinematic bounce distance}
\label{sec:bounce}

A non-singular bounce must satisfy the following conditions:
\begin{equation}\label{bounce-condition}
 H(t_b)=0
 \quad\land\quad
 \dot H(t_b)>0\,.
\end{equation}
In canonical Einstein-scalar theory:
\begin{equation}
    \dot H=-\frac{\sigmadot^2}{2\mpl^2}\le 0\,,
\end{equation}
so a bounce requires NEC violation, modified gravity, ghost-condensate/Galileon/Horndeski/DHOST-type dynamics, loop effects or additional sectors \cite{Battefeld:2014uga,Brandenberger:2016vhg,Ilyas:2020zcb,Zhu:2021whu}. A universal perturbative Lyth-like formula cannot be assigned to this phase. Nevertheless, its field-space distance is kinematically well defined:
\begin{equation}
 \db=\frac1{\mpl}\int_{t_-}^{t_+}\sigmadot\,dt\,.
\end{equation}
Let us also define:
\begin{equation}
 \Delta t_b=t_+-t_-
 \quad\land\quad
 \Delta H_b=H(t_+)-H(t_-)>0\,,
\end{equation}
and a phenomenological \textit{effective bounce coefficient}:
\begin{equation}
 \dot H=\frac{\Xi(t)}{2\mpl^2}\sigmadot^2\,,
\end{equation}
with:
\begin{equation}
 \Xi_{\rm eff}\equiv\frac{2\mpl^2\,\Delta H_b}{\int_{t_-}^{t_+}\sigmadot^2dt}\,.
 \label{eq:xi-eff}
\end{equation}
In the case of usual canonical evolution:
\begin{equation}
    \Xi=-1\,.
\end{equation}
so that, a successful bounce requires an effective positive contribution. By introducing:
\begin{equation}
 I_1\equiv\int_{t_-}^{t_+}\sigmadot\,dt
 \quad\land\quad
 I_2\equiv\int_{t_-}^{t_+}\sigmadot^{2}\,dt\,,
\end{equation}
and the shape factor:
\begin{equation}
 S_b\equiv\frac{I_1^2}{\Delta t_b\,I_2}
 \quad\land\quad
 0<S_b\le 1\,,
 \label{eq:Sb}
\end{equation}
where the upper limit follows from the Cauchy-Schwarz inequality:
\begin{equation}\label{Cauchy-Schwarz}
 \left(\int_{t_-}^{t_+}\dot\sigma\,dt\right)^2\le\left(\int_{t_-}^{t_+}\dot\sigma^2\,dt\right)\left(\int_{t_-}^{t_+}dt\right)=I_2\,\Delta t_b\,.
\end{equation}
Using the fact that:
\begin{equation}
    I_2=2\mpl^2\frac{\Delta H_b}{\Xi_{\rm eff}}
\end{equation}
we get:
\begin{equation}
 \db^2=\frac{I_1^2}{\mpl^2}=S_b\,\Delta t_b\frac{I_2}{\mpl^2}=\frac{2S_b\,\Delta H_b\,\Delta t_b}{\Xi_{\rm eff}}\,,
\end{equation}
so that:
\begin{equation}
 \db=\left(\frac{2S_b\,\Delta H_b\,\Delta t_b}{\Xi_{\rm eff}}\right)^{1/2}\,.
 \label{eq:db-general}
\end{equation}
For a symmetric bounce from $H_-=-H_B$ to $H_+=+H_B$, the parameters are given by:
\begin{equation}
 \Delta H_b=2H_B
 \quad\land\quad
 \tau_b\equiv H_B\Delta t_b\,,
\end{equation}
which yields:
\begin{equation}
 \db=2\sqrt{\frac{S_b\,\tau_b}{\Xi_{\rm eff}}}\,.
 \label{eq:db-symmetric}
\end{equation}
If only a distance $d_{{\rm rem},b}$ remains for the bounce, then:
\begin{equation}
 \tau_b\le \frac{\Xi_{\rm eff}}{4S_b}d_{{\rm rem},b}^2\,.
 \label{eq:taub-bound}
\end{equation}
Thus, a small-field bounce is short in units of the bounce scale unless:
\begin{equation}
    \Xi_{\rm eff}\gg1
    \quad\land\quad
    S_b\ll 1\,,
\end{equation}
or the bounce itself consumes a distance of the Planckian order.

\subsection{Mechanism-resolved bounce sectors}
\label{sec:mechanism-bounce}

The parametrization in \eqref{eq:db-symmetric} should not be interpreted as a universal dynamical law for all non-singular bounces. It is just a kinematic identity for a scalar field-mediated effective bounces if the quantities $S_b$, $\tau_b$ and $\Xi_{\rm eff}$ can be defined. Different microscopic bounce mechanisms assign different meanings to the bounce contribution in the total field-space budget. From that reason, it is useful to write:
\begin{equation}
 \db\mapsto d_b^{(\mathcal B)}\,,
 \label{eq:db-mechanism-def}
\end{equation}
where $\mathcal B$ denotes the bounce mechanism. The distance criterion now becomes:
\begin{equation}
 \frac{\sqrt{2\epsilon_{\rm ek}}}{\left(\epsilon_{\rm ek}-1\right)}\Nsm+\dconv+d_b^{(\mathcal B)}+\dpost\le d_{\rm allowed}^{(\mathcal B)}\,.
 \label{eq:master-mechanism}
\end{equation}
Such a form is more general than \eqref{eq:db-symmetric}. The latter is recovered when the bounce is mediated by the same scalar sector whose invariant distance is being counted. An explicit smooth scalar-mediated toy realization of this statement is given in Appendix~\ref{app:toy-scalar-bounce}. A detailed description of the specific realizations of cosmological models with bounce/cyclic behavior discussed below can be found in Table~\ref{tab:bounce-mechanisms}.

\begin{table*}[htbp]
\caption{Mechanism-dependent interpretation of the bounce contribution $d_b^{(\mathcal B)}$. In scalar-mediated bounces it is a genuine field-space distance. In quantum-gravity-driven or thin-matching descriptions it is more accurately interpreted as an effective cost, a cutoff/curvature-control criterion, or a constraint on transfer functions.}
\label{tab:bounce-mechanisms}
\begin{ruledtabular}
\begin{tabular}{p{0.18\textwidth}p{0.25\textwidth}p{0.25\textwidth}p{0.23\textwidth}}
Mechanism $\mathcal B$ & Origin of bounce & Meaning of $d_b^{(\mathcal B)}$ & Additional consistency criteria \\
\hline
Ghost condensate/Galileon & Controlled effective NEC violation by higher-derivative scalar dynamics \cite{Arkani-Hamed:2003pdi,Cai:2012va,Battarra:2014tga,Ijjas:2016tpn} & Genuine distance of the bounce scalar, well approximated by \eqref{eq:db-symmetric} when the scalar sector dominates & No ghost instability, controlled gradient instability and EFT cutoff above the bounce scale \\
DHOST/Horndeski-type bounce & Degenerate higher-order scalar-tensor dynamics \cite{Ilyas:2020zcb,Zhu:2021whu} & Scalar distance plus scalar-tensor EFT-control cost; $\Xi_{\rm eff}$ is an effective description rather than a canonical NEC parameter & Degeneracy criteria, $Q_s>0$, $Q_T>0$, stable sound speeds and controlled transfer functions \\
Matter/quintom bounce & Multifield or effective NEC violation in a matter-like contracting background \cite{Novello:2008ra,Brandenberger:2012zb,Brandenberger:2016vhg} & Multifield distance through the NEC-violating sector, $M_{\rm Pl}^{-1}\int \sqrt{G_{IJ}\dot\phi^I\dot\phi^J}\,dt$ & Ghost/gradient instabilities and anisotropy growth without an ekpyrotic smoothing phase \\
$f(R)$/modified gravity & Curvature-sector dynamics, often expressible through a scalaron in the Einstein frame \cite{Sotiriou:2006hs,Sotiriou:2008rp} & Scalaron or gravitational-sector distance, e.g. $d_b^{(f(R))}=\sqrt{3/2}\,|\Delta\ln f_R|$ if the map is regular & Frame regularity, absence of extra instabilities and cutoff control \\
Galilean Genesis/conformal Galileon & Higher-derivative scalar EFT transition or bounce \cite{Qiu:2011cy,Ijjas:2016tpn} & Scalar displacement plus higher-derivative EFT-control cost, $d_b^{({\rm Gen})}=d_{\rm scalar}+d_{\rm EFT}$ & Strong coupling and gradient stability \\
Loop quantum cosmology & Effective quantum-geometry correction to the Friedmann equation \cite{Ashtekar:2011ni,Cai:2014zga} & Not necessarily a scalar distance; often better treated as a curvature or quantum-gravity validity cost & $\rho\le\rho_c$, validity of effective equations and a specified perturbation prescription \\
S-brane or thin matching & Localized transition or matching surface \cite{Gutperle:2002ai,Kounnas:2011gz,Brandenberger:2013zea,Brandenberger:2020eyf,Brandenberger:2020tcr} & Often $d_b^{({\rm S})}\simeq0$ in scalar field space, with the cost shifted to matching and transfer functions & Well-defined matching conditions and stable scalar/tensor transfer across the transition \\
CCC conformal crossover & Conformal identification of the future infinity of one aeon with the Big Bang of the next \cite{Penrose:1980ge,Penrose:2006zz,Penrose:2010,Penrose:2010zz,Tod:2013rsa,Meissner:2015ita,Markwell:2022noq,Meissner:2025gsa} & Not a metric bounce; $d_b$ is not naturally defined unless a dynamical conformal representative is specified & Choice of conformal factor, crossover physics and observational status
\end{tabular}
\end{ruledtabular}
\end{table*}
\subsubsection{Ghost condensate and Galileon}
For a ghost condensate or Galileon bounces, the scalar field itself is responsible for the NEC-violating phase \cite{Arkani-Hamed:2003pdi,Battarra:2014tga,Ijjas:2016tpn}. In that case $d_b^{({\rm GC/Gal})}$ is naturally identified with the invariant scalar distance:
\begin{equation}
 d_b^{({\rm GC/Gal})}=\frac{1}{\mpl}\int_{t_-}^{t_+}\sigmadot_b\,dt\,,
 \label{eq:db-gc}
\end{equation}
and the estimate \eqref{eq:db-symmetric} directly constrains the duration of the bounce. This distance criterion must still be supplemented by stability and cutoff conditions for the higher-derivative scalar sector.

\subsubsection{DHOST and Horndeski}
For DHOST\footnote{Degenerate Higher-Order Scalar-Tensor.} or Horndeski-type bounces, the relation between $\dot H$ and the scalar velocity is modified by higher-derivative scalar-tensor operators. A useful schematic decomposition is:
\begin{equation}
 d_b^{({\rm DHOST})}=d_{\sigma,b}+d_{\rm hd}\,,
 \label{eq:db-dhost}
\end{equation}
where $d_{\sigma,b}$ is the direct scalar distance and $d_{\rm hd}$ represents the cost of keeping the higher-derivative scalar-tensor EFT under control. A complete model must compute the kinetic coefficients and sound speeds of scalar and tensor perturbations, as well as the transfer functions, all of which must be done through the bounce. This is important because generic, non-singular cosmologies in Horndeski/Galileon-type theories are subject to no-go and instability constraints \cite{Kobayashi:2016xpl,Cai:2016thi}.

\subsubsection{LQC}
For a loop quantum cosmology-type effective bounce, the modified Friedmann equation is often written as \cite{Ashtekar:2011ni,Cai:2014zga}:
\begin{equation}
 H^2=\frac{\rho}{3\mpl^2}\left(1-\frac{\rho}{\rho_c}\right)\,,
 \label{eq:lqc-friedmann}
\end{equation}
with the effective Raychaudhuri equation of the form:
\begin{equation}
 \dot H=-\frac{\left(\rho+p\right)}{2\mpl^2}\left(1-\frac{2\rho}{\rho_c}\right)\,.
 \label{eq:lqc-raychaudhuri}
\end{equation}
In the case of:
\begin{equation}
    \rho+p=\sigmadot^2\,,
\end{equation}
comparison with \eqref{eq:xi-eff} suggests the effective identification:
\begin{equation}
 \Xi_{\rm LQC}(t)=\frac{2\rho(t)}{\rho_c}-1\,.
 \label{eq:xi-lqc}
\end{equation}
Thus, the LQC contribution should be treated not simply as a field-space cost, but as a curvature-control or quantum gravity validity criterion:
\begin{equation}
 d_b^{({\rm LQC})}=d_{\sigma,b}+d_{\rm QG}\,.
 \label{eq:db-lqc}
\end{equation}

\subsubsection{Matter bounce and quintom}
Matter bounce or quintom-type realizations are closer to scalar-mediated bounces \cite{Novello:2008ra,Brandenberger:2012zb}, but the relevant distance must be computed in the full target space of the fields responsible for the NEC-violating stage, namely:
\begin{equation}
 d_b^{({\rm q})}=\frac{1}{\mpl}\int_{\rm b}\sqrt{G_{IJ}\dot\phi^I\dot\phi^J}\,dt\,.
 \label{eq:db-quintom}
\end{equation}
This class is especially sensitive to ghost \cite{Carroll:2003st,Cline:2003gs,Cai:2009zp} or gradient instabilities \cite{Libanov:2016kfc,Kobayashi:2016xpl} and to the growth of anisotropy \cite{Cai:2013vm,Qiu:2013eoa} if a matter-like contraction is not preceded or supplemented by ekpyrotic smoothing.

\subsubsection{$f(R)$ theories}
For $f(R)$ or related modified gravity bounces, the distance may reside in the gravitational sector rather than in a matter scalar \cite{Sotiriou:2006hs,Sotiriou:2008rp}. In the scalar-tensor representation of $f(R)$ gravity, with:
\begin{equation}
    f_{R}\equiv\frac{df(R)}{dR}\,,
\end{equation}
the Einstein-frame scalaron takes the form:
\begin{equation}
 \chi=\sqrt{\frac{3}{2}}\,\mpl \ln{\left(f_R\right)}\,,
 \label{eq:scalaron}
\end{equation}
and the corresponding contribution is:
\begin{equation}
 d_b^{(f(R))}=\sqrt{\frac{3}{2}}\,\left|\Delta \ln{\left(f_R\right)}\right|\,,
 \label{eq:db-fr}
\end{equation}
assuming the scalar-tensor map is regular over the relevant domain, the distance-budget language can be applied to curvature-sector bounces. This shows that the distance-budget language can be used for curvature sector bounces, but only after the field representation is specified and stability is verified.

\subsubsection{Galilean Genesis and conformal Galileon}
Galilean Genesis or conformal Galileon transitions \cite{Qiu:2011cy,Ijjas:2016tpn} should also be distinguished from a canonical scalar bounce scenarios. They are naturally represented by:
\begin{equation}
 d_b^{({\rm Gen})}=d_{\rm scalar}^{({\rm Gen})}+d_{\rm EFT}^{({\rm Gen})}\,.
 \label{eq:db-genesis}
\end{equation}
The second term summarizes the requirements for strong coupling, cutoff, and higher-derivative stability. This notation is schematic by design; a specific Genesis or conformal Galileon model must directly compute the kinetic coefficients, sound speeds, and transfer functions.

\subsubsection{CCC}
Conformal cyclic cosmology\footnote{For a more detailed explanation of Penrose's motivations for the CCC and his hypothesis that gravity could modify quantum mechanics ('gravitization' of QM) rather than simply being quantized in the conventional sense, see \cite{Penrose:1996cv,Penrose:2014nha,Penrose:2014vok,Penrose:2017wxd}.} (CCC) \cite{Penrose:1980ge,Penrose:2006zz,Penrose:2010,Penrose:2010zz,Tod:2013rsa,Meissner:2015ita,Markwell:2022noq,Meissner:2025gsa} is a specific type of cyclic cosmological model, but it does not involve a bounce in the usual metric sense. Rather than a local transition characterized by the bounce condition, the future conformal infinity of one aeon is identified with the conformally rescaled Big Bang of the next aeon. Consequently, the variable $d_b$ used above is not suitable for CCC. To interpret distance budgets, one must first specify a dynamical conformal factor or a scalar representative of the conformal rescaling. For a logarithmic conformal variable:
\begin{equation}
    \chi=\alpha\,\mpl\ln{\Omega}\,,
\end{equation}
one may define an effective crossover distance:
\begin{equation}
 d_{\rm cross}^{({\rm CCC})}=\alpha\left|\Delta\ln\Omega\right|\,.
 \label{eq:dccc-cross}
\end{equation}
However, since the conformal factor in CCC is part of the matching structure rather than a propagating scalar field, this quantity is model-dependent and should not be included in the universal bounce-distance term. CCC therefore aligns with the current framework only as a conformal crossover extension:
\begin{equation}
 d_{\rm cosmic}^{({\rm CCC})}=d_{\rm aeon}+d_{\rm cross}^{({\rm CCC})}\,,
 \label{eq:dccc-cosmic}
\end{equation}
rather than as a non-singular scalar bounce.

\subsubsection{S-brane and thin-matching}
In an S-brane or thin-matching description \cite{Gutperle:2002ai,Kounnas:2011gz,Brandenberger:2013zea,Brandenberger:2020eyf,Brandenberger:2020tcr}, the scalar field-space distance consumed during the transition may be negligible:
\begin{equation}
 d_b^{({\rm S})}\simeq 0\,.
 \label{eq:db-sbrane}
\end{equation}
However, the cost of the bounce is shifted to the requirement that the matching conditions generate stable and observationally viable scalar and tensor transfer functions. The distinction between the two is important because the distance budget constrains the background trajectory, while the observable tensor and scalar amplitudes depend on model-specific transfer functions.

This mechanism-resolved interpretation leads to the generalized effective distance:
\begin{equation}
 d_{\rm eff}^{(\mathcal B)}=d_{\rm allowed}^{(\mathcal B)}-\dconv-d_b^{(\mathcal B)}-\dpost\,.
 \label{eq:deff-mechanism}
\end{equation}
Whenever:
\begin{equation}
    d_{\rm eff}^{(\mathcal B)}>0\,,
\end{equation}
the smoothing criterion becomes:
\begin{equation}
 \epsilon_{\rm ek}\ge\left[\frac{\Nsm}{\sqrt2\,d_{\rm eff}^{(\mathcal B)}}
 +\sqrt{1+\frac{\Nsm^2}{2\left[d_{\rm eff}^{(\mathcal B)}\right]^2}}
 \right]^2\,.
 \label{eq:eps-mechanism}
\end{equation}
If $d_{\rm eff}^{(\mathcal B)}\le 0$, the chosen bounce mechanism and conversion history are incompatible with the adopted finite trajectory length criterion. This does not constitute a model-independent exclusion based on geodesic distance swampland reasoning alone. The mechanism-dependent fingerprints in the distance-budget for each of the discussed models one can find in Table~\ref{tab:bounce-fingerprints}.

\begin{table*}[htbp]
\caption{Mechanism-dependent fingerprints in the distance-budget framework. The entries are qualitative because each mechanism requires its own perturbation and stability analysis. The table is intentionally mechanism-resolved: the bounce contribution is a genuine scalar distance only for scalar-mediated bounces, while in quantum gravity or matching descriptions the dominant constraint may instead be EFT validity or perturbation transfer.}
\label{tab:bounce-fingerprints}
\begin{ruledtabular}
\begin{tabular}{p{0.22\textwidth}p{0.14\textwidth}p{0.28\textwidth}p{0.25\textwidth}}
Mechanism & Is $d_b$ large? & Main observational channel & Main theoretical risk \\
\hline
Ghost condensate/Galileon & possible & non-Gaussianity, scalar transfer, tensor suppression & cutoff control and gradient stability \\
DHOST/Horndeski & model-dependent & scalar tilt, tensor-to-scalar ratio, tensor transfer & degeneracy conditions and perturbative stability \\
Matter/quintom & possible & scalar transfer and isocurvature & ghost/gradient instabilities and anisotropy growth \\
$f(R)$/modified gravity & model-dependent & scalaron dynamics, tensor transfer & frame regularity and extra degrees of freedom \\
Galilean Genesis & model-dependent & scalar tilt, non-Gaussianity, strong-coupling signatures & strong coupling and gradient stability \\
Loop quantum cosmology & not necessarily & transfer through the high-curvature regime & validity of effective equations and perturbation prescription \\
S-brane/matching & usually small & tensor tilt and matching-induced transfer & UV completion and matching prescription \\
CCC crossover & not a bounce & conformal-crossover signatures, possible CMB anomalies & conformal-factor choice and crossover physics
\end{tabular}
\end{ruledtabular}
\end{table*}

\section{Master distance budget inequality}
\label{sec:master}

Combining \eqref{eq:dek-constant}, \eqref{eq:dconv-bound}, \eqref{eq:db-symmetric} and \eqref{eq:dallowed}, one obtains:
\begin{equation}
 \frac{\sqrt{2\epsilon_{\rm ek}}}{\epsilon_{\rm ek}-1}\Nsm
 +\frac{\Thetareq}{\mpl\kappa_{g,{\rm max}}}+2\sqrt{\frac{S_b\tau_b}{\Xi_{\rm eff}}}+\dpost\le \dallowed\,.
 \label{eq:master}
\end{equation}
By defining the effective distance available for a smoothing:
\begin{equation}
 \deff\equiv \dallowed-\frac{\Thetareq}{\mpl\kappa_{g,{\rm max}}}
 -2\sqrt{\frac{S_b\tau_b}{\Xi_{\rm eff}}}-\dpost\,.
 \label{eq:deff}
\end{equation}
the viable small-distance completion requires $\deff>0$. For $\deff>0$ one may obtain:
\begin{equation}
 \epsilon_{\rm ek}\ge\left[\frac{\Nsm}{\sqrt2\,\deff}+\sqrt{1+\frac{\Nsm^2}{2\deff^2}}\right]^2\,.
 \label{eq:eps-master-exact}
\end{equation}
Moreover, for $\Nsm\gg\deff$ it yields:
\begin{equation}
 \epsilon_{\rm ek}\gtrsim \frac{2\Nsm^2}{\deff^2}\,.
 \label{eq:eps-master-approx}
\end{equation}
The BKL anisotropy requirement gives the additional exact constraint of the form:
\begin{equation}
 \epsilon_{\rm ek}\ge\left[\frac{\ABKL}{2\sqrt2\,\deff}+\sqrt{3+\frac{\ABKL^2}{8\deff^2}}\right]^2\,.
 \label{eq:eps-BKL-master-exact}
\end{equation}
Therefore, in the large $\epsilon$ regime, the result of combining smoothing and isotropization is:
\begin{equation}
 \epsilon_{\rm ek}\gtrsim\max\left[\frac{2\Nsm^2}{\deff^2},\frac{\ABKL^2}{2\deff^2}\right]\,.
 \label{eq:eps-master-max}
\end{equation}
The relationships \eqref{eq:master-mechanism}-\eqref{eq:eps-mechanism} and their scalar-mediated specialization \eqref{eq:master}-\eqref{eq:eps-master-max} are the central proposed results of this study. They make precise the statement that \textit{ekpyrotic smoothing, anisotropy suppression, entropy conversion, and bounce dynamics are not independent from the point of view of field-space control}.

\section{Applications and diagnostics}
\label{sec:benchmarks}

Whether the preceding inequalities are useful depends on their ability to establish transparent restrictions on model parameters. This section collects several benchmark applications that can be used to diagnose concrete ekpyrotic or bouncing constructions.

\subsection{Exponential ekpyrotic potential}

For the single-field exponential potential:
\begin{equation}
 V(\phi)=-V_0e^{-c\phi/\mpl}
 \quad\land\quad
 \epsilon_{\rm ek}=\frac{c^2}{2}\,,
 \label{eq:benchmark-exp}
\end{equation}
the relation \eqref{eq:eps-master-exact} immediately turns into a conditional lower limit on the slope\footnote{For a standard scaling-solution analysis of exponential potentials, see \cite{Copeland:1997et}.}. We can deduce that in the large $\epsilon$ limit:
\begin{equation}
 c_{\rm min}\simeq\frac{2\Nsm}{\deff}\,.
 \label{eq:cmin-benchmark}
\end{equation}
For $\Nsm=60$, representative values are shown in Table~\ref{tab:exp-benchmark}. The table illustrates the central message of the distance budget: \textit{once conversion, bounce and post-bounce phases consume part of the available field distance, the required ekpyrotic slope becomes substantially steeper}.

\begin{figure}[htbp]
 \centering
 \includegraphics[width=\columnwidth]{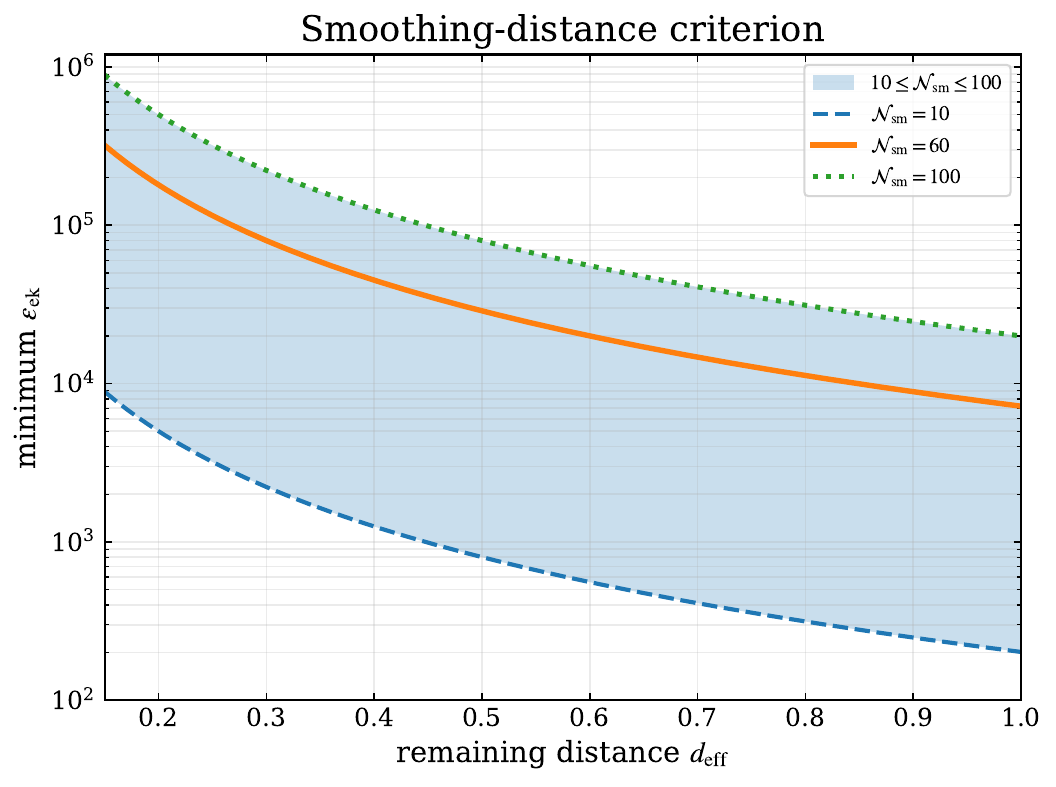}
 \caption{Minimum ekpyrotic fast-roll parameter as a function of the remaining field-space distance \(\deff\). The shaded band covers \(10\le\Nsm\le100\), with representative curves highlighted. Decreasing \(\deff\), for example by allocating field distance to entropy conversion, late-time motion, or the bounce, rapidly strengthens the conditional lower limit on \(\epsilon_{\rm ek}\).}
 \label{fig:epsilon-deff}
\end{figure}

This dependence is visualized in Fig.~\ref{fig:epsilon-deff}. 
The plot shows that the conditional lower limit on \(\epsilon_{\rm ek}\) increases rapidly as the remaining distance \(\deff\) decreases. Thus, even moderate distance costs from entropy conversion, the bounce, or post-bounce evolution can push the ekpyrotic phase into the ultra-fast-roll regime.

\begin{figure}[htbp]
\centering
\includegraphics[width=\columnwidth]{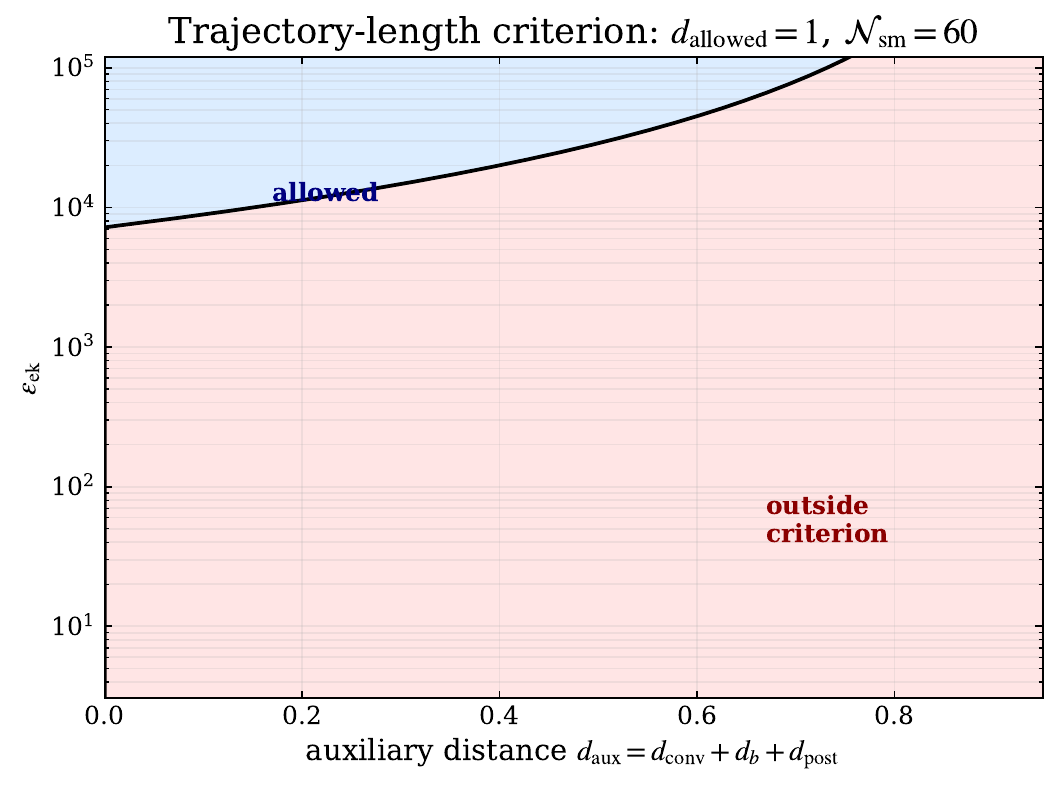}
\caption{Allowed and disfavored regions under the adopted trajectory length criterion in the auxiliary-distance plane for \(d_{\rm allowed}=1\) and \(\Nsm=60\). Here \(d_{\rm aux}=d_{\rm conv}+d_b+d_{\rm post}\), so increasing \(d_{\rm aux}\) leaves less distance for ekpyrotic smoothing and therefore increases the required \(\epsilon_{\rm ek}\).}
\label{fig:allowed-daux}
\end{figure}

Equivalently, one may display the same constraint in terms of the 
auxiliary distance. For fixed \(d_{\rm allowed}\), increasing \(d_{\rm aux}\) decreases:
\begin{equation}
    \deff=d_{\rm allowed}-d_{\rm aux}\,,
\end{equation}
moving the model toward the region outside the adopted trajectory length criterion shown in Fig.~\ref{fig:allowed-daux}.

\begin{table}[htbp]
 \caption{Benchmark conditional lower limits for the exponential potential \(V(\phi)=-V_0e^{-c\phi/\mpl}\), assuming \(\Nsm=60\). The asymptotic estimates use \(\epsilon_{\rm min}\simeq2\Nsm^2/\deff^2\) and \(c_{\rm min}\simeq2\Nsm/\deff\).}
 \label{tab:exp-benchmark}
 \centering
 \begin{tabular}{ccc}
 \toprule
 \(\deff\) & \(\epsilon_{\rm ek,min}\) & \(c_{\rm min}\) \\
 \midrule
 1.00 & \(7.2\times10^3\) & 120 \\
 0.75 & \(1.28\times10^4\) & 160 \\
 0.50 & \(2.88\times10^4\) & 240 \\
 0.25 & \(1.15\times10^5\) & 480 \\
 \bottomrule
 \end{tabular}

\end{table}

\subsection{Application to cyclic ekpyrotic potentials}
\label{sec:cyclic-potentials}

The distance-budget criterion can also be applied directly to potentials used as simple $4D$ descriptions of cyclic ekpyrotic cosmology. The distance budget constrains not only the asymptotic steepness parameter, but also the field interval on which the ekpyrotic phase can consistently occur. We write the dimensionless field variable as:
\begin{equation}
 \varphi\equiv\frac{\phi}{\mpl}\,,
\end{equation}
and define the local steepness parameter:
\begin{equation}
 \lambda(\varphi)\equiv -\frac{1}{V}\frac{dV}{d\varphi}=-\mpl\frac{V_{,\phi}}{V}\,.
 \label{eq:lambda-def-benchmark}
\end{equation}
For an approximately scaling ekpyrotic regime:
\begin{equation}
 \epsilon_{\rm pot}(\varphi)\simeq\frac{\lambda^2(\varphi)}{2}\,,
 \label{eq:epsilon-pot-lambda}
\end{equation}
so the distance budget criterion requires:
\begin{equation}
 \min_{\varphi\in I_{\rm ek}}\left|\lambda(\varphi)\right|\gtrsim \lambda_{\rm req}
 \quad\land\quad
 \lambda_{\rm req}\equiv\sqrt{2\epsilon_{\rm ek,min}}\,.
 \label{eq:lambda-req}
\end{equation}
In the large \(\epsilon\) limit, \eqref{eq:eps-master-approx} produces:
\begin{equation}
 \lambda_{\rm req}\simeq\frac{2\Nsm}{\deff}\,.
 \label{eq:lambda-req-asym}
\end{equation}

A first cyclic potential is of the form \cite{Steinhardt:2002ih,Steinhardt:2001st}:
\begin{equation}
 V_{1}(\varphi)=V_0\left(1-e^{-c\varphi}\right)\,.
 \label{eq:cyclic-potential-one}
\end{equation}
It has a positive plateau as \(\varphi\to\infty\) and a negative steep ekpyrotic branch for \(\varphi<0\). On the negative branch:
\begin{equation}
 \lambda_1(\varphi)=\frac{c}{1-e^{c\varphi}}
 \quad\land\quad
 \varphi<0\,.
 \label{eq:lambda-one}
\end{equation}
Hence \(\lambda_1\to c\) far on the negative branch and \(\lambda_1\to\infty\) as the zero of the potential is approached from below.

A second standard cyclic toy potential is \cite{Lehners:2008vx,Lehners:2010fy}:
\begin{equation}
 V_{2}(\varphi)=V_0\left(e^{b\varphi}-e^{-c\varphi}\right)
 \quad\land\quad
 b\ll1
 \quad\land\quad
 c\gg1\,,
 \label{eq:cyclic-potential-two}
\end{equation}
which captures both a shallow positive dark energy branch and a steep negative ekpyrotic branch\footnote{For more details on the stability of this class of models, see \cite{Postolak:2025qmv}.}. In complete cyclic models this type of potential is often supplemented by a cutoff function that turns off the steep negative part near the end of ekpyrosis \cite{Steinhardt:2001st,Steinhardt:2002ih,Lehners:2008vx}. For \(\varphi<0\):
\begin{equation}
 \lambda_{2}(\varphi)=\frac{b e^{b\varphi}+c e^{-c\varphi}}{e^{-c\varphi}-e^{b\varphi}}\,.
 \label{eq:lambda-two}
\end{equation}
Again \(\lambda_{2}\to c\) far on the negative branch, while on the positive branch \(|\lambda_2|\to b\), giving a slowly rolling dark energy sector if \(b\ll1\).

For both cyclic potentials, a robust ekpyrotic branch over an interval \(I_{\rm ek}\) is obtained for:
\begin{equation}
 c\gtrsim\lambda_{\rm req}\simeq\frac{2\Nsm}{\deff}\,.
 \label{eq:cyclic-c-bound}
\end{equation}
For \(\Nsm=60\), the corresponding steepness requirement is summarized in Table~\ref{tab:cyclic-lambda-benchmark}. In the cyclic potentials \(V_1\) and \(V_2\), this table should be read as a limit on the local branch: if the asymptotic value \(c\) is below \(\lambda_{\rm req}\), the criterion can only be met close to the zero of the potential, where the long scaling approximation is typically lost.

\begin{table}[htbp]
 \caption{Distance-budget requirement for the local steepness \(\lambda_{\rm req}\simeq2\Nsm/\deff\), assuming \(\Nsm=60\). For the cyclic potentials \(V_1\) and \(V_2\), a robust scaling branch requires the asymptotic parameter \(c\) to be at least of order \(\lambda_{\rm req}\).}
 \label{tab:cyclic-lambda-benchmark}
 \centering
 \begin{tabular}{ccc}
 \toprule
 \(\deff\) & \(\lambda_{\rm req}\) & \(\epsilon_{\rm ek,min}\simeq\lambda_{\rm req}^2/2\) \\
 \midrule
 1.00 & 120 & \(7.2\times10^3\) \\
 0.75 & 160 & \(1.28\times10^4\) \\
 0.50 & 240 & \(2.88\times10^4\) \\
 0.25 & 480 & \(1.15\times10^5\) \\
 \bottomrule
 \end{tabular}
\end{table}

In the case of \(c<\lambda_{\rm req}\), the formal inequality can be satisfied only close to the zero of the potential, where \(\lambda\) becomes large. This is not the usual scaling ekpyrotic regime and generally yields only a restricted field interval. More explicitly, for the scalar field potential \(V_1\) and \(\lambda_{\rm req}>c\), the condition \(\lambda_1\ge\lambda_{\rm req}\) requires:
\begin{equation}
 \varphi\ge\varphi^{(1)}_{\lambda}\equiv\frac{1}{c}\ln\left(1-\frac{c}{\lambda_{\rm req}}\right)
 \quad\land\quad
 \varphi<0\,.
 \label{eq:phi-lambda-one}
\end{equation}
For \(V_2\), the analogous condition gives:
\begin{equation}
 \varphi\ge\varphi^{(2)}_{\lambda}\equiv\frac{1}{b+c}\ln\left(
 \frac{\lambda_{\rm req}-c}{\lambda_{\rm req}+b}\right)
 \quad\land\quad
 \varphi<0\,.
 \label{eq:phi-lambda-two}
\end{equation}
Thus, the distance budget not only constrains the asymptotic steepness, $c$, but also restricts where on the cyclic potential the ekpyrotic phase may occur if $c$ is smaller than the required value. This is a useful diagnostic tool because successful cyclic models require a combination of a shallow dark energy branch, controlled entropy production, and a sufficiently steep negative branch within one scalar trajectory.

\subsection{Smoothing versus anisotropy suppression}
\begin{figure*}[htbp]
 \centering
 \includegraphics[width=1\textwidth]{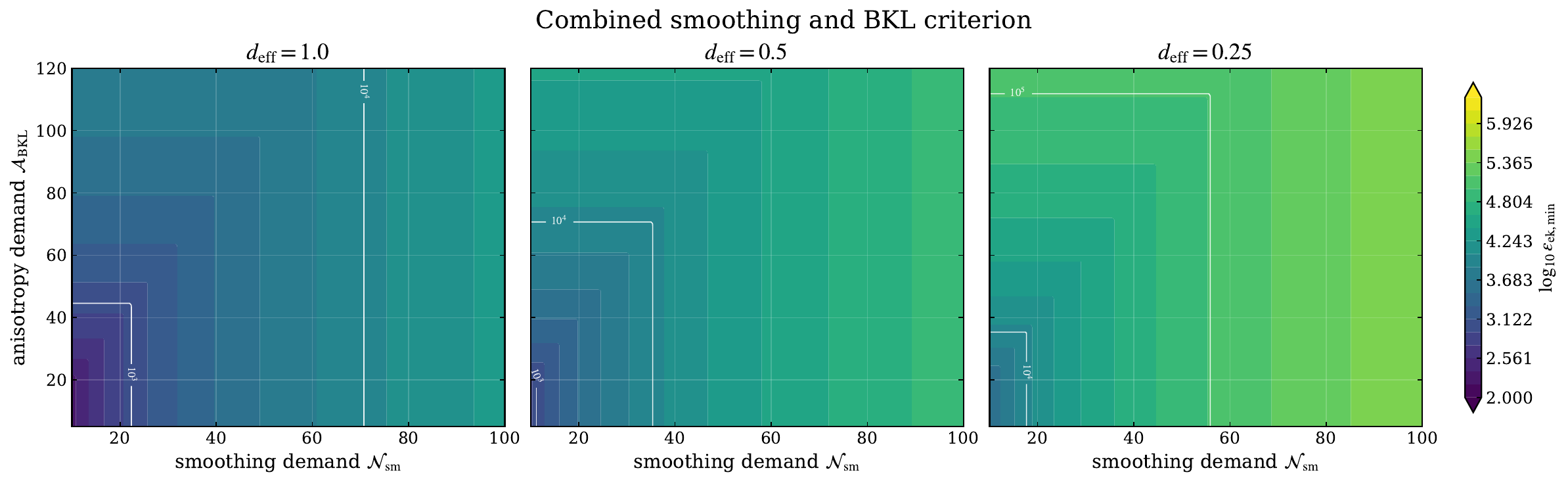}
 \caption{Combined smoothing and BKL anisotropy criterion shown as a two-parameter diagnostic. The panels scan the plane $(\Nsm,\ABKL)$ for representative remaining distances $\deff=\left\{1,0.5,0.25\right\}$. The color scale gives \(\log_{10}\epsilon_{\rm ek,min}\), illustrating how both the smoothing demand and the initial shear burden tighten the fast-roll requirement.}
 \label{fig:BKL-smoothing}
\end{figure*}

The BKL criterion introduces an additional physical target. For a fixed $\deff$, the minimum $\epsilon_{\rm ek}$ is determined by whichever is larger of the smoothing and anisotropy requirements:
\begin{equation}
 \epsilon_{\rm ek,min}\simeq\max\left[\frac{2\Nsm^2}{\deff^2},\frac{\ABKL^2}{2\deff^2}\right]\,.
 \label{eq:benchmark-max}
\end{equation}
This is illustrated in Fig.~\ref{fig:BKL-smoothing}, which displays the combined criterion in the $(\Nsm,\ABKL)$ plane for various remaining distance values. This plot should be interpreted as a \textit{diagnostic map rather than a prediction}. The $\ABKL$ value depends on the assumed initial shear and the duration of the ekpyrotic phase.

\subsection{Conversion and bounce windows}
\begin{figure}[htbp]
 \centering
 \includegraphics[width=\columnwidth]{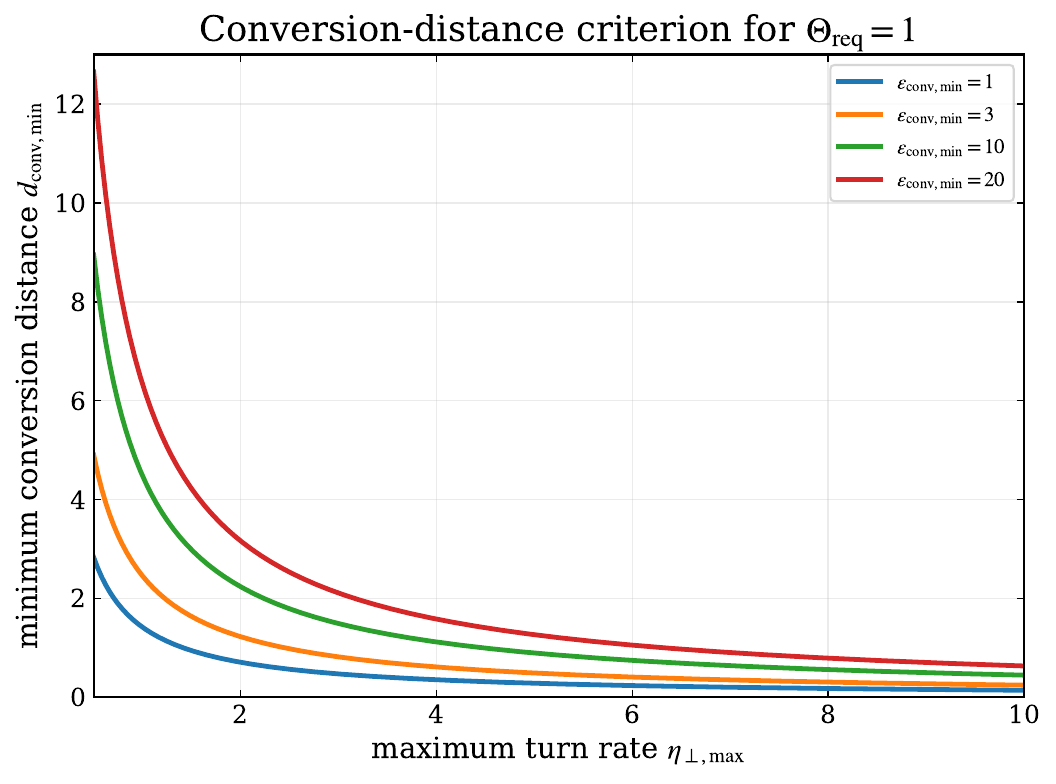}
 \caption{Illustrative turn-rate conversion limit. The curves show the minimum conversion distance implied by \(d_{\rm conv}\gtrsim
\sqrt{2\epsilon_{\rm conv,min}}\Theta_{\rm req}/\eta_{\perp,\max}\)
for \(\Theta_{\rm req}=1\). Isocurvature and non-Gaussianity constraints
can be incorporated by replacing \(\eta_{\perp,\max}\) with the appropriate
model-dependent observational limit.}
 \label{fig:conversion-window}
\end{figure}

The conversion and bounce sectors can be presented within the same framework of budgetary constraints. Fig.~\ref{fig:conversion-window} illustrates the minimum conversion distance obtained via the phenomenological isocurvature and non-Gaussianity criteria of Sec.~\ref{sec:conversion}.

\begin{figure}[htbp]
 \centering
 \includegraphics[width=\columnwidth]{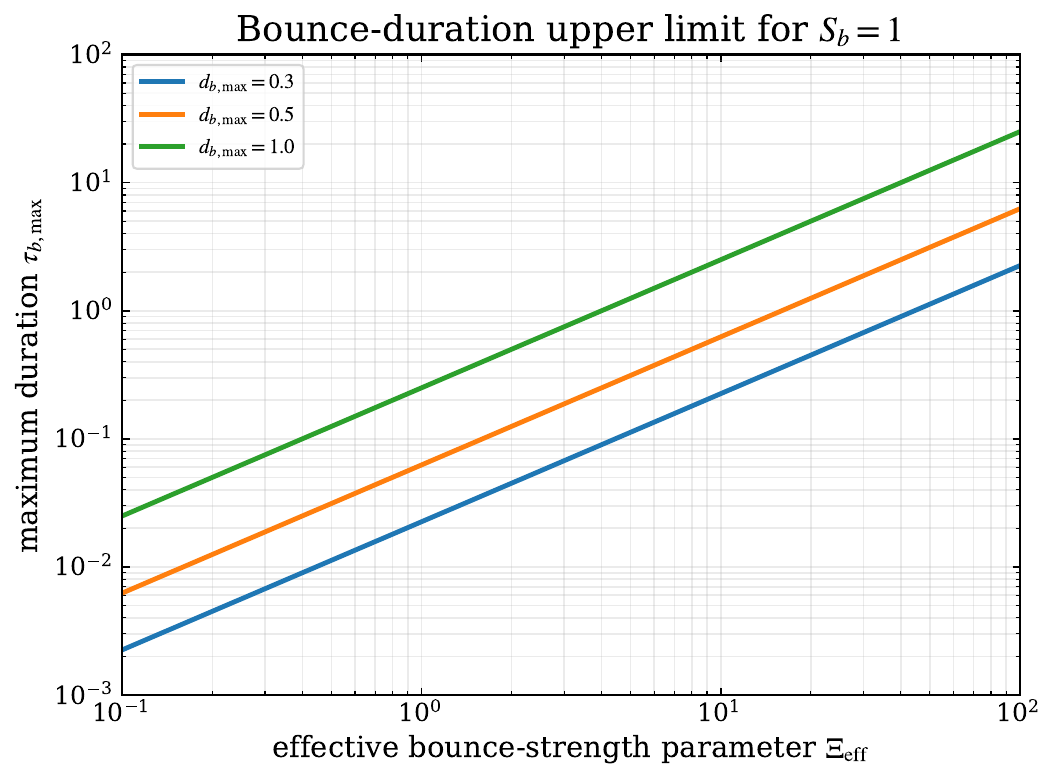}
 \caption{Maximum symmetric-bounce duration \(\tau_{b,{\rm max}}\) as a function of the effective bounce strength \(\Xi_{\rm eff}\), for several values of the remaining bounce distance. For \(S_b=1\), a small remaining distance forces the bounce to be short unless \(\Xi_{\rm eff}\) is large.}
 \label{fig:bounce-window}
\end{figure}

Meanwhile, Fig.~\ref{fig:bounce-window} demonstrates the largest possible bounce interval within the symmetric-bounce parametrization. Together, these plots more explicitly demonstrate the qualitative statement that a small-field completion requires conversion that is both efficient and observationally acceptable, as well as a bounce that is either short or strongly dominated by modified gravity.

\subsection{Field-space curvature diagnostic}

\begin{figure}[htbp]
 \centering
 \includegraphics[width=\columnwidth]{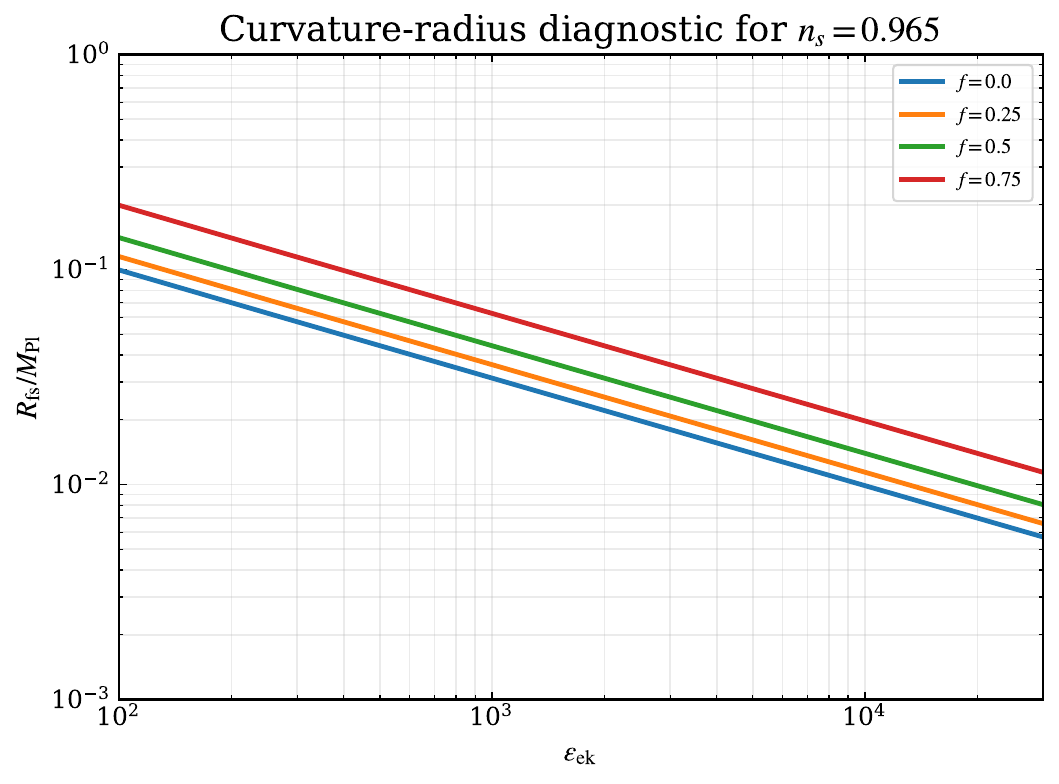}
 \caption{Indicative field-space curvature radius required by the entropy-sector tilt condition for \(n_s=0.965\). The parameter \(f\) denotes the fraction of the required entropy mass supplied by \(V_{;NN}\). The curves include \(f=\left\{0,0.25,0.5,0.75\right\}\). Smaller \(f\) corresponds to a larger geometrical contribution and hence a smaller curvature radius.}
 \label{fig:curvature-radius}
\end{figure}

For entropic mechanisms, the observed scalar tilt constrains the entropy mass and therefore the field-space curvature contribution. Fig.~\ref{fig:curvature-radius} shows the curvature radius implied by the geometrical contribution for \(n_s=0.965\), when a fraction \(f\) of the required entropy mass is supplied by the transverse potential curvature. The rapid decrease of \(R_{\rm fs}\) with \(\epsilon_{\rm ek}\) illustrates why small-distance ultra-fast-roll ekpyrosis tends to require either a highly curved target space or a strongly tachyonic transverse potential.

\section{Entropy tilt and field-space curvature}
\label{sec:entropy-tilt}

The distance budget can be connected to the observed scalar tilt in entropic ekpyrotic models. Now, let us consider a two-field trajectory with entropy perturbation $Q_s$. In the limit in which the entropy fluctuation is approximately decoupled during its generation, the canonically normalized variable:
\begin{equation}
    v_s=a\,Q_s
\end{equation}
obeys the equation of motion:
\begin{equation}
 v_s''+\left(k^2+a^2m_s^2-\frac{a''}{a}\right)v_s=0\,,
 \label{eq:vs-eq}
\end{equation}
where primes denote derivatives with respect to the conformal time $\tau$. For a constant $\epsilon$ parameter during the contraction phase:
\begin{equation}
 a(\tau)\propto(-\tau)^q,
 \quad\land\quad
 q=\frac{1}{\left(\epsilon-1\right)}\,,
\end{equation}
so that:
\begin{equation}
 \frac{a''}{a}=\frac{q\left(q-1\right)}{\tau^2}
 \quad\land\quad
 a^{2}H^{2}=\frac{q^2}{\tau^2}\,.
\end{equation}
Introducing the dimensionless effective entropy mass parameter:
\begin{equation}
    \mu_{s}\equiv\frac{m_{s}^2}{H^2}\,,
\end{equation}
the relation \eqref{eq:vs-eq} becomes:
\begin{equation}
 v''_{s}+\left[k^2+\frac{\mu_{s}\,q^{2}-q\left(q-1\right)}{\tau^2}\right]v_{s}=0\,.
\end{equation}
Comparison with the standard Bessel form for a canonical
cosmological perturbation mode \cite{Mukhanov:1990me,Brandenberger:2003vk,Lehners:2008vx}:
\begin{equation}
 v''_s+\left[k^2-\frac{\nu_s^2-\frac{1}{4}}{\tau^2}\right]v_{s}=0
\end{equation}
gives:
\begin{equation}
 \nu_{s}^2=\frac{1}{4}+\frac{2-\epsilon-\mu_s}{\left(\epsilon-1\right)^2}\,,
\end{equation}
with:
\begin{equation}\label{ns-nus}
 n_{s}-1=3-2\nu_{s}
 \quad\land\quad
 \nu_{s}=\frac{4-n_{s}}{2}\,,
\end{equation}
one obtains
\begin{equation}
 \mu_{s}\equiv\frac{m_{s}^{2}}{H^2}=2-\epsilon-\left(\epsilon-1\right)^{2}\left[\left(\frac{4-n_{s}}{2}\right)^2-\frac{1}{4}\right]\,.
 \label{eq:mu-ns}
\end{equation}
For exact scale invariance, $n_s=1$, this reduces to
\begin{equation}
 \frac{m_s^2}{H^2}=-2\epsilon^2+3\epsilon.
 \label{eq:ms-scale-invariant}
\end{equation}
Thus, scale-invariant entropy fluctuations in an ultra-ekpyrotic phase require a large negative effective entropy mass.

For a curved field-space manifold, the entropy mass contains a geometrical contribution\footnote{We use the Riemann-tensor convention for which negative sectional curvature \(K_{\rm fs}=R_{TNTN}<0\) gives a tachyonic geometrical contribution.} \cite{Gordon:2000hv,Gong:2011uw,Renaux-Petel:2015mga}:
\begin{equation}
 m_{s}^{2}=V_{;NN}+3\Omega^2+\sigmadot^2 R_{TNTN}\,,
 \label{eq:ms-curved}
\end{equation}
where:
\begin{equation}
 R_{TNTN}=R_{IJKL}T^{I}N^{J}T^{K}N^{L}\equiv K_{\rm fs}
\end{equation}
is the sectional curvature in the adiabatic-entropic plane. Since $\sigmadot^2=2\epsilon\mpl^2H^2$ and $\eta_\perp=\Omega/|H|$, therefore:
\begin{equation}
 \frac{m_s^2}{H^2}=\frac{V_{;NN}}{H^2}+3\eta_\perp^2+2\epsilon\mpl^2K_{\rm fs}\,.
\end{equation}
Combining this with \eqref{eq:mu-ns}, we obtain:
\begin{equation}
 \mpl^2K_{\rm fs}=\frac{2-\epsilon-\left(\epsilon-1\right)^2\left[\left(\frac{4-n_s}{2}\right)^2-\frac{1}{4}\right]-\frac{V_{;NN}}{H^2}-3\eta_{\perp}^{2}}{2\epsilon}\,.
 \label{eq:Kfs-master}
\end{equation}
For $n_s=1$ and negligible turning during the entropy generation:
\begin{equation}
 \mpl^2K_{\rm fs}\simeq -\epsilon+\frac{3}{2}-\frac{1}{2\epsilon}\frac{V_{;NN}}{H^2}\,.
 \label{eq:Kfs-scale-invariant}
\end{equation}
If the transverse potential curvature does not supply the tachyonic mass, then:
\begin{equation}
 K_{\rm fs}\sim -\frac{\epsilon}{\mpl^2}
\end{equation}
and the curvature radius takes the form of:
\begin{equation}
    R_{\rm fs}\sim \frac{\mpl}{\sqrt{\epsilon}}\,.
\end{equation}
In a consequence, the small-field distance budget can imply a highly curved scalar manifold. For $\epsilon\sim 10^4$, the curvature radius becomes:
\begin{equation}
    R_{\rm fs}\sim10^{-2}\mpl\,.
\end{equation}
It is also useful to decompose the entropy mass requirement into the fractional contributions:
\begin{equation}
 f_{V}\equiv\frac{\frac{V_{;NN}}{H^2}}{\mu_s}
 \quad\land\quad
 f_{\Omega}\equiv\frac{3\eta_{\perp}^2}{\mu_s}
 \quad\land\quad
 f_{K}\equiv\frac{2\epsilon\,\mpl^{2}\,K_{\rm fs}}{\mu_s}\,.
 \label{eq:entropy-mass-fractions}
\end{equation}
In addition, these fractions obey the following condition:
\begin{equation}
 f_V+f_{\Omega}+f_K=1\,.
 \label{eq:entropy-fraction-sum}
\end{equation}
For \(\epsilon\gg1\), the required entropy mass is generically of order:
\begin{equation}
    \mu_{s}\sim -2\epsilon^{2}\,.
\end{equation}
Therefore, in a small-distance ekpyrotic model, a significant amount of entropy must be allocated to transverse potential curvature, negative field-space curvature or turning. This makes equation \eqref{eq:entropy-mass-fractions} a useful diagnostic tool for determining how the observed tilt is dynamically supported.

The observed running constrains not only the instantaneous value of \(\mu_s\), but also its flow along the ekpyrotic smoothing trajectory. Using \eqref{ns-nus} and \eqref{eq:mu-ns} gives:
\begin{equation}
 \mu_{s}=2-\epsilon-\left(\epsilon-1\right)^2\left(\nu_{s}^2-\frac{1}{4}\right)\,.
\end{equation}
Differentiating with respect to \(\Nsm\), and writing:
\begin{equation}
    \epsilon'\equiv\frac{d\epsilon}{d\Nsm}\,,
\end{equation}
gives:
\begin{equation}
 \frac{d\mu_s}{d\Nsm}=-\epsilon'-2\left(\epsilon-1\right)\,\epsilon'\left(\nu_{s}^{2}-\frac{1}{4}\right)+\left(\epsilon-1\right)^{2}\,\nu_{s}\,\alpha_{s}\,,
 \label{eq:mu-flow}
\end{equation}
up to the sign convention relating \(\ln{k}\) to \(\Nsm\). Here $\alpha_{s}$ denotes the \textit{running of the scalar spectral index}:
\begin{equation}
    \alpha_{s}\equiv\frac{d n_{s}}{d \Nsm}\simeq\frac{d n_{s}}{d \ln{k}}\,.
\end{equation}
Moreover, the dimensionless entropy-mass flow parameter:
\begin{equation}
 \mathcal{F}_{s}\equiv\left|\frac{d\ln{\left|\mu_{s}\right|}}{d\Nsm}\right|
 \label{eq:entropy-flow-diagnostic}
\end{equation}
measures how rapidly the tachyonic entropy sector, transverse mass, or field-space curvature must vary. A small \(\deff\) increases the required \(\epsilon_{\rm ek}\), and consequently makes a given measured running more demanding for the microscopic entropy sector.

\section{Connections to present and future data}
\label{sec:data}
The proposed distance budget does not provide a universal prediction for the tensor-to-scalar ratio, $r$. Rather, it provides a framework for translating observational constraints into limitations on smoothing, conversion, and bounce parameters.

\subsection{Scalar tilt, running and entropy sector allocation}
Planck found a red scalar tilt of the value:
\begin{equation}
    n_{s}\simeq 0.965\,,
\end{equation}
and there is no strict evidence for a running in the corcondance model \cite{Planck:2018vyg}. Furthermore, bounds were set on residual isocurvature contributions at the percent level, and the exact bound depended on the isocurvature mode and the correlation choice \cite{Planck:2018jri}. Additionally, we did not detect significant primordial non-Gaussianity within local, equilateral or orthogonal templates \cite{Planck:2019kim}. According to the present framework, the tilt determines the instantaneous entropy mass via equation \eqref{eq:mu-ns}, while the running determines the entropy-mass flow via equations \eqref{eq:mu-flow} and \eqref{eq:entropy-flow-diagnostic}. The observational information is therefore translated into the pair:
\begin{equation}
 \left(n_{s},\alpha_{s}\right)\quad\mapsto\quad\left(\mu_{s},\mathcal{F}_{s}\right)\,.
 \label{eq:tilt-running-map}
\end{equation}
This map is useful because the distance budget fixes a conditional lower limit on \(\epsilon_{\rm ek}\). Once \(\epsilon_{\rm ek}\) is large, \eqref{eq:entropy-mass-fractions} tells us how the required tachyonic entropy mass must be allocated between potential curvature, turning and target-space geometry. Lehners found that small-field scale-free ekpyrosis can naturally generate running of the order \cite{Lehners:2018vgi}:
\begin{equation}
    \alpha_{s}\sim -10^{-3}\,.
\end{equation}
The phase-resolved budget sharpens this statement: conversion, bounce, and late-time distance costs reduce \(\deff\), raise \(\epsilon_{\rm ek}\), and thereby make any nonzero running a constraint on how rapidly the entropy sector may vary.

\subsection{Non-Gaussianity, residual isocurvature and conversion viability}
Sharp entropy-to-curvature conversion is not observationally free. Planck finds no primordial non-Gaussianity signal and obtains stringent bounds on the local, equilateral and orthogonal templates \cite{Planck:2019kim}. Ekpyrotic conversion mechanisms can generate significant non-Gaussianity, depending on the conversion dynamics \cite{Lehners:2010fy}. The conversion distance criterion:
\begin{equation}
 \dconv\ge\frac{\Thetareq}{\mpl\,\kappa_{g,{\rm max}}}
\end{equation}
shows that a sub-Planckian conversion range requires either a large curvature of the trajectory or a large dimensionless turn rate. These are precisely the features that can enhance non-Gaussianity in entropic ekpyrotic conversion and, more generally, in rapidly turning multifield systems \cite{Lehners:2007wc,Lehners:2008vx,Lehners:2010fy,Iarygina:2023msy}.

A useful measure of conversion sharpness is the \textit{average sharpness of the entropy-to-curvature conversion turn}:
\begin{equation}
 \mathcal S_{\rm turn}\equiv\frac{\Theta_{\rm req}}{\Delta n_{\rm conv}}=\frac{\Theta_{\rm req}\sqrt{2\epsilon_{\rm conv}}}{\dconv}\,.
 \label{eq:turn-sharpness}
\end{equation}
Model-dependent non-Gaussianity constraints can then be represented as:
\begin{equation}
    \mathcal S_{\rm turn}\lesssim\eta_{\perp,{\rm NG}}\,,
\end{equation}
giving the conditional lower limit \(d_{\rm conv}^{\rm NG}\) already introduced in \eqref{eq:dconv-NG}.

In the similar manner, using a transfer matrix formalism, one could write \cite{Wands:2002bn,Gordon:2000hv,Malik:2008im}:
\begin{equation}
 \mathcal{R}_{f}=\mathcal{R}_{i}+T_{\mathcal{RS}}\mathcal{S}_{i}
 \quad\land\quad
 \mathcal{S}_{f}=T_{\mathcal{SS}}\mathcal{S}_{i}\,,
\end{equation}
which shows that the residual isocurvature constraints impose efficient conversion and give the conditional lower limit \(d_{\rm conv}^{\rm iso}\) in \eqref{eq:dconv-isocurvature}. The conversion sector is therefore viable only if the observationally required minimum conversion distance fits inside the remaining budget. By defining:
\begin{equation}
 d_{\rm conv}^{\rm min}=\max\left(d_{\rm conv}^{\rm iso},d_{\rm conv}^{\rm NG}\right)\,,
 \label{eq:dconv-min}
\end{equation}
a necessary criterion for a controlled conversion phase is:
\begin{equation}
 d_{\rm conv}^{\rm min}<d_{\rm allowed}^{(\mathcal B)}-d_{\rm ek}-d_b^{(\mathcal B)}-d_{\rm post}\,.
 \label{eq:conversion-viability}
\end{equation}
This inequality constitutes a direct distance-budget consistency test: conversion scenarios that require either too much bending or too sharp a turn are incompatible with a prescribed finite field-space range.

\subsection{Tensor modes, transfer gains and stochastic backgrounds}
Current CMB observations strongly constrain the tensor-to-scalar ratio. For example, BICEP/Keck in combination with Planck and WMAP yields:
\begin{equation}
    r_{0.05}<0.036
\end{equation}
at the 95\% confidence level \cite{BICEP:2021xfz}. Future experiments, such as CMB-S4 and LiteBIRD, aim to improve sensitivity by one to two orders of magnitude\footnote{LiteBIRD targets a total uncertainty \(\delta r<10^{-3}\) for a fiducial \(r=0\) model \cite{LiteBIRD:2022cnt}.} \cite{CMB-S4:2016ple,LiteBIRD:2022cnt}. The detection of nearly scale-invariant tensor spectrum would put significant pressure on minimal ekpyrotic models, but not on all modified bounce or sourced-tensor scenarios.

For a mechanism \(\mathcal B\), one can define phenomenological scalar and
tensor transfer functions through the bounce or matching surface
\cite{Deruelle:1995kd,Brandenberger:2013zea,Brandenberger:2020tcr,Zhu:2021whu}:
\begin{equation}
 \mathcal G_T^{(\mathcal B)}(k)\equiv\left|\mathcal T_T^{(\mathcal B)}(k)\right|^2
 \quad\land\quad
 \mathcal G_{\mathcal R}^{(\mathcal B)}(k)\equiv\left|\mathcal T_{\mathcal R}^{(\mathcal B)}(k)\right|^2\,,
 \label{eq:transfer-gains}
\end{equation}
so that:
\begin{equation}
 r_f(k)=r_{\rm in}(k)\frac{\mathcal G_T^{(\mathcal B)}(k)}{\mathcal G_{\mathcal R}^{(\mathcal B)}(k)}\,.
 \label{eq:r-transfer-gain}
\end{equation}
The current upper bound on \(r\) therefore constrains the relative transfer gain:
\begin{equation}
    \frac{\mathcal G_T^{(\mathcal B)}(k_*)}{\mathcal G_{\mathcal R}^{(\mathcal B)}(k_*)}<\frac{r_{\rm max}}{r_{\rm in}(k_*)}\,.
\label{eq:relative-gain-bound}
\end{equation}
Therefore, a bounce mechanism must either begin with a suppressed input tensor amplitude or increase the transfer of scalars sufficiently to stay below the CMB limits.

Pulsar timing arrays have reported some evidence for a nanohertz stochastic
GWs background
\cite{NANOGrav:2023gor,EPTA:2023fyk,Reardon:2023gzh,Xu:2023wog}. The source may be astrophysical, but cosmological sources remain an important target for model building. For a stochastic background, we can write:
\begin{equation}
 \Omega_{\rm GW,0}(f)=\Omega_{\rm GW,in}(f)\mathcal G_T^{(\mathcal B)}(f)
 \mathcal R_{\rm red}(f)\,,
 \label{eq:OmegaGW-gain}
\end{equation}
where \(\mathcal R_{\rm red}\) summarizes the ordinary redshifting and thermal history factors. A useful phenomenological diagnostic of the bounce response is:
\begin{equation}
 \mathcal B_T(f)\equiv\frac{d\ln\mathcal G_T^{(\mathcal B)}(f)}{d\ln f}\,.
 \label{eq:bounce-spectral-response}
\end{equation}
A sharp or matching-like transition may produce a localized feature in \(\mathcal B_T\), whereas a smoother scalar-mediated bounce is expected to give a broader transfer response. This produces a compact way to connect \((d_b^{(\mathcal B)},\tau_b,\Xi_{\rm eff})\) to multi-band stochastic-background constraints.

\begin{figure*}[htbp]
 \centering
 \includegraphics[width=0.85\textwidth]{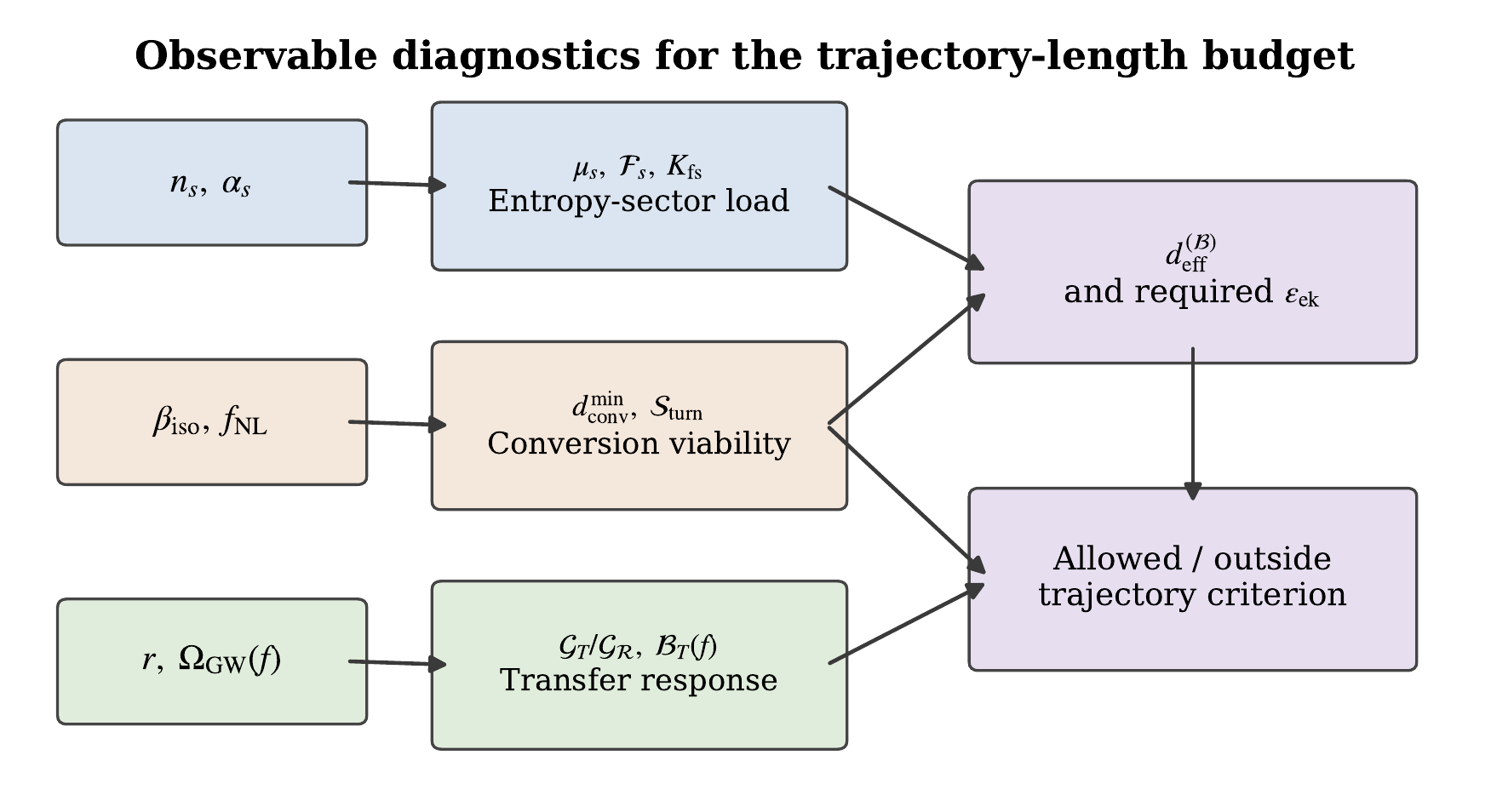}
 \caption{Observable-to-budget diagnostic map. The scalar tilt and running constrain the entropy-sector load; isocurvature and non-Gaussianity constrain conversion viability; tensors and stochastic backgrounds constrain transfer response. These inputs feed into the effective trajectory length budget and the regions allowed or disfavored under the adopted criterion.}
 \label{fig:observable-diagnostics-map}
\end{figure*}

The observable diagnostics for the field-space distance budget are summarized in Table~\ref{tab:observable-diagnostics} and Fig.~\ref{fig:observable-diagnostics-map}.

\begin{table*}[htbp]
\caption{Observable diagnostics of the distance budget framework. The entries are phenomenological maps rather than model-independent numerical predictions; each concrete bounce or conversion mechanism must supply its own perturbation transfer functions. These diagnostic links are summarized in Fig.~\ref{fig:observable-diagnostics-map}.}
\label{tab:observable-diagnostics}
\begin{ruledtabular}
\begin{tabular}{p{0.16\textwidth}p{0.28\textwidth}p{0.45\textwidth}}
Observable & Distance-budget quantity & Interpretation \\
\hline
\(n_s\) & \(\mu_s=m_s^2/H^2\), \(K_{\rm fs}\) & Fixes the required entropy mass and the burden placed on potential curvature, turning or field-space curvature \\
\(\alpha_s\) & \(\mathcal F_s=|d\ln|\mu_s|/d\Nsm|\) & Measures how rapidly the entropy sector must vary during the smoothing phase \\
\(\beta_{\rm iso}\) & \(d_{\rm conv}^{\rm iso}\) & Requires sufficiently efficient entropy-to-curvature conversion \\
\(f_{\rm NL}\) & \(\mathcal S_{\rm turn}\), \(d_{\rm conv}^{\rm NG}\) & Limits how sharp the conversion turn can be \\
\(r\) & \(\mathcal G_T/\mathcal G_{\mathcal R}\) & Constrains relative tensor/scalar transfer through the bounce and conversion sectors \\
\(\Omega_{\rm GW}(f)\) & \(\mathcal B_T(f)\), \(d_b^{(\mathcal B)}\), \(\tau_b\) & Constrains the tensor transfer response of the bounce or crossover mechanism \\
\end{tabular}
\end{ruledtabular}
\end{table*}

\subsection{DESI-motivated CPL distance and cyclic slope hierarchy}
\label{sec:desi-cpl-distance}
Recent DESI DR2 BAO results \cite{DESI:2025zgx,DESI:2025fii}, when combined with CMB and supernovae data, have reported a preference for evolving dark energy in $\omega_{0}\omega_{a}$CDM model, although the interpretation remains under active scrutiny and may depend on data combinations and possible systematics. The present use of DESI DR2 is therefore not a claim that scalar field dark energy has been established. Rather, DESI-motivated CPL reconstructions provide a useful phenomenological input for asking how much late-time field-space distance would be consumed if the observed trend were interpreted as scalar sector dynamics.

\begin{figure}[htbp]
 \centering
 \includegraphics[width=\columnwidth]{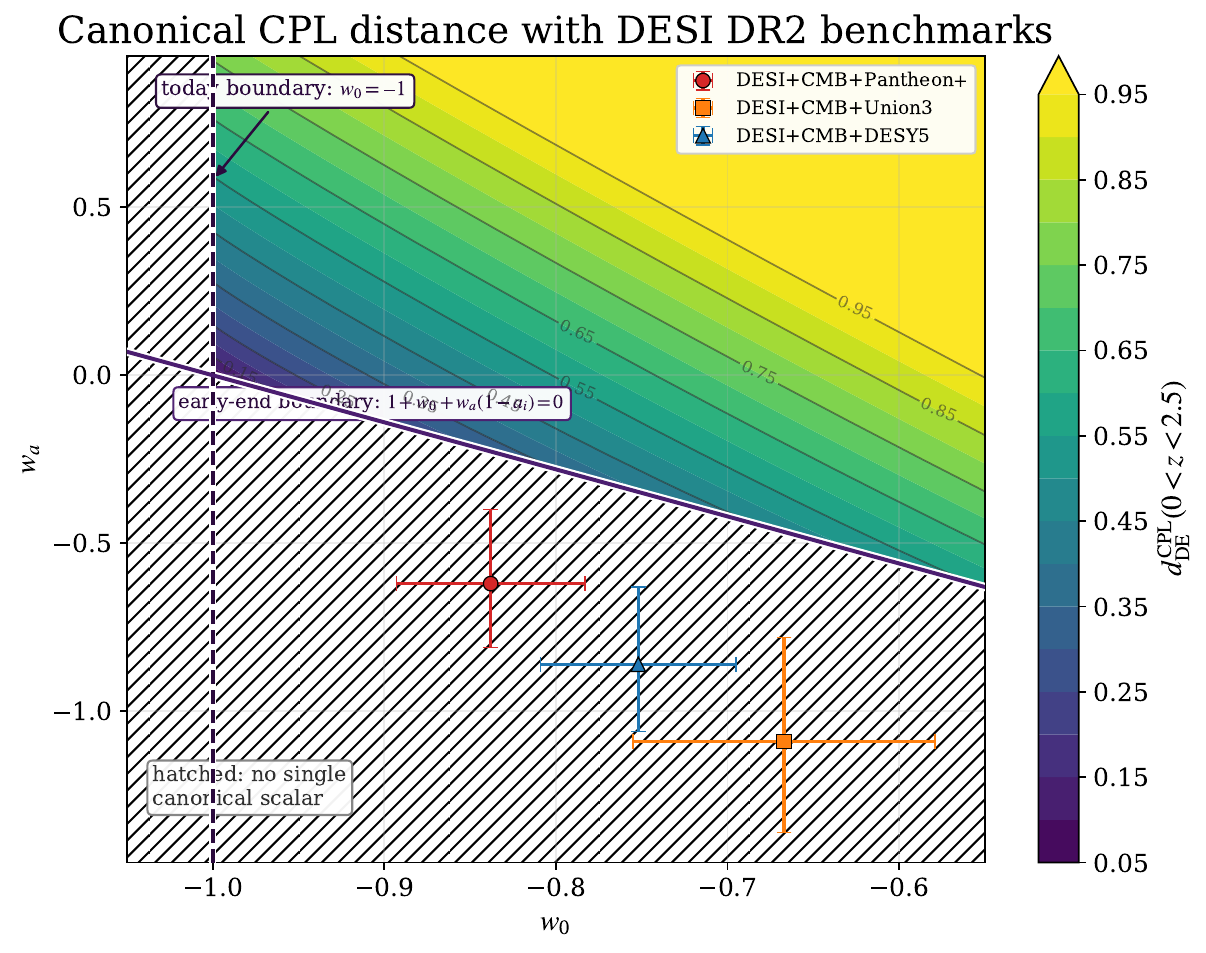}
 \caption{Canonical late-time scalar-field distance $d_{\rm DE}^{\rm CPL}$ for the CPL equation of state over $0<z<2.5$, using $\Omega_{m0}=0.315$, with the canonical-domain boundaries and the quoted DESI DR2 benchmark results superposed as $1\sigma$ error bars. The hatched region corresponds to parameter values for which $1+\omega(a)<0$ at some point within the interval, resulting in a single canonical quintessence interpretation failing. The boundary guides show that $\omega_0=-1$ and that $1+\omega_{0}+\omega_{a}(1-a_i)=0$, where $a_i=1/(1+2.5)\simeq 0.286$.}
 \label{fig:CPL-distance}
\end{figure}

For canonical quintessence with $\omega_\phi>-1$ \cite{Copeland:2006wr}:
\begin{equation}
 \dot\phi^2=\left(1+\omega_{\phi}\right)\,\rho_{\phi}
 \quad\land\quad
 \rho_{\phi}=3\mpl^{2}\,H^{2}\,\Omega_\phi\,,
\end{equation}
so that:
\begin{equation}
 d_{\rm DE}\left(a_1,a_2\right)=\int_{a_1}^{a_2}d\ln{a}\,\sqrt{3\,\Omega_{\phi}(a)\left[1+\omega_\phi(a)\right]}\,.
 \label{eq:dDE}
\end{equation}
For the Chevallier-Polarski-Linder (CPL) parametrization \cite{Chevallier:2000qy,Linder:2002et}:
\begin{equation}
 \omega(a)=\omega_{0}+\omega_{a}\left(1-a\right)\,,
 \label{eq:CPL-w}
\end{equation}
and, in the case of a spatially flat Universe filled by matter (baryonic and DM) and dark energy:
\begin{equation}
 E^2(a)=\Omega_{m0}\,a^{-3}+\left(1-\Omega_{m0}\right)\,a^{-3(1+\omega_{0}+\omega_{a})}e^{-3\omega_{a}(1-a)}\,,
 \label{eq:CPL-E2}
\end{equation}
with:
\begin{equation}
 \Omega_{\rm DE}(a)=\frac{(1-\Omega_{m0})\,a^{-3(1+\omega_{0}+\omega_{a})}e^{-3\omega_{a}(1-a)}}{E^2(a)}\,.
 \label{eq:CPL-OmegaDE}
\end{equation}
Therefore, the canonical CPL distance can be written as:
\begin{equation}
 d_{\rm DE}^{\rm CPL}(a_i,1)=\int_{a_i}^{1}d\ln{a}\,\sqrt{3\Omega_{\rm DE}(a)\left[1+\omega_{0}+\omega_{a}(1-a)\right]}\,.
 \label{eq:dDE-CPL}
\end{equation}
This expression is real only on the canonical side of the phantom divide line. Since the CPL function is linear in $a$, a canonical single-field interpretation over $a\in[a_i,1]$ requires:
\begin{equation}
 1+\omega_{0}\ge 0
 \quad\land\quad
 1+\omega_{0}+\omega_{a}(1-a_i)\ge 0\,.
 \label{eq:CPL-canonical-region}
\end{equation}
The phantom-crossing scale factor is:
\begin{equation}
 a_{\times}=1+\frac{1+\omega_{0}}{\omega_a}
 \quad\land\quad
 z_{\times}=\frac{1}{a_{\times}}-1\,,
 \label{eq:CPL-crossing}
\end{equation}
when $a_{\times}$ lies in the observed interval. If the preferred reconstruction crosses $\omega=-1$, the late-time sector cannot be represented by a single canonical scalar distance; it must instead be interpreted as a quintom, multifield, non-minimally coupled or modified gravity distance. Figure~\ref{fig:CPL-distance} shows the canonical distance map with three DESI DR2 benchmark determinations quoted in the text, shown as $1\sigma$ error bars for the combinations DESI+CMB+Pantheon+, DESI+CMB+Union3, and DESI+CMB+DESY5.

\begin{figure}[htbp]
 \centering
 \includegraphics[width=\columnwidth]{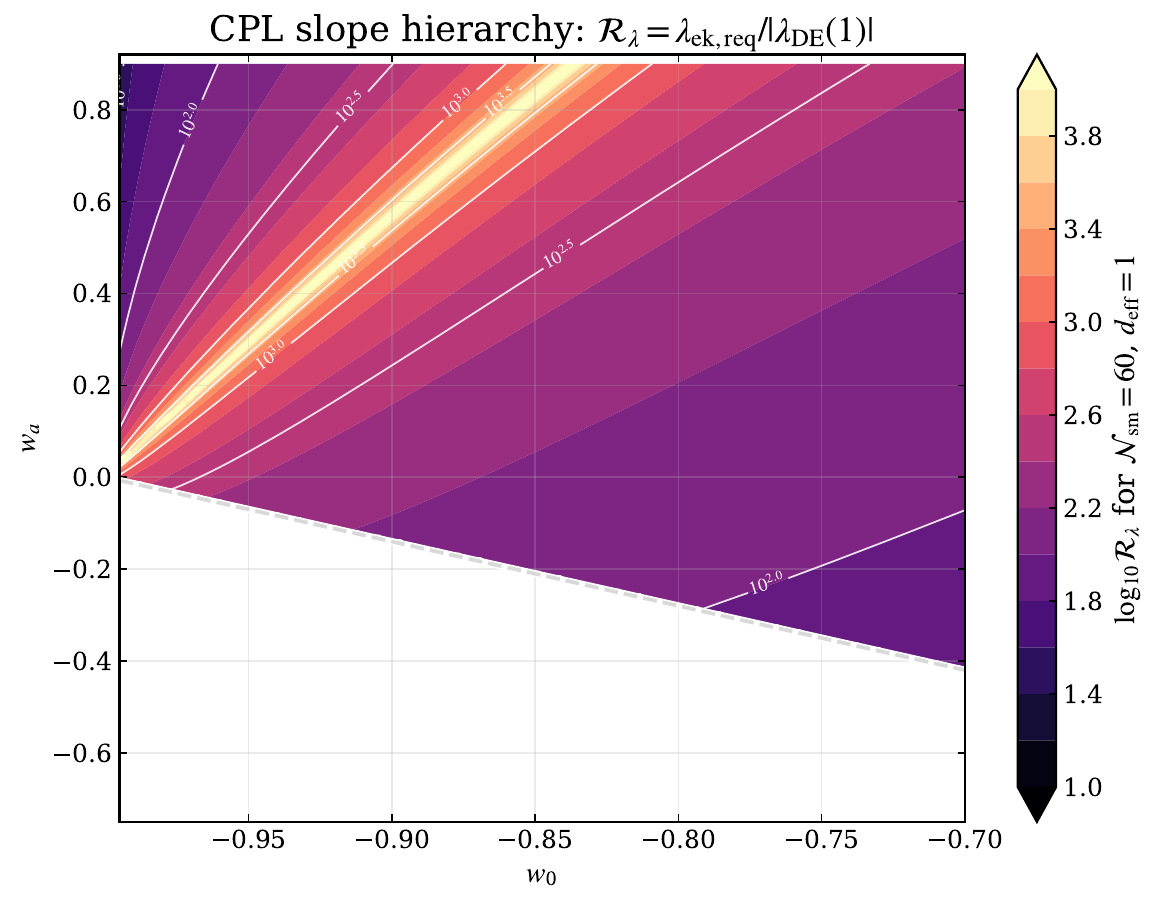}
 \caption{CPL slope-hierarchy diagnostic in the canonical region. The contours show the hierarchy \(\mathcal R_\lambda=\lambda_{\rm ek,req}/|\lambda_{\rm DE}(1)|\), using \(\Nsm=60\) and \(\deff=1\). Large values indicate that a cyclic scalar sector must interpolate between a shallow late-time branch and an ultra-steep ekpyrotic branch.}
 \label{fig:CPL-slope-hierarchy}
\end{figure}

The same CPL reconstruction also gives a useful local slope diagnostic. For a canonical quintessence:
\begin{equation}
 V(a)=\frac{1-\omega(a)}{2}\rho_{\rm DE}(a)
\end{equation}
and:
\begin{equation}
  \frac{1}{\mpl}\frac{d\phi}{d\ln a}=\pm\sqrt{3\Omega_{\rm DE}(a)[1+\omega(a)]}\,.
\end{equation}
Using the fact that:
\begin{equation}
    \frac{d\ln\rho_{\rm DE}}{d\ln{a}}=-3\left[1+\omega(a)\right]
\end{equation}
and:
\begin{equation}
    \frac{d\omega(a)}{d\ln{a}}=-a\,\omega_{a}\,,
\end{equation}
the magnitude of the reconstructed CPL slope is of the following form:
\begin{equation}
 \left|\lambda_{\rm DE}^{\rm CPL}(a)\right|=\left|\frac{3[1+\omega(a)]-\frac{a\,\omega_a}{1-\omega(a)}}{\sqrt{3\Omega_{\rm DE}(a)\left[1+\omega(a)\right]}}\right|\,,
 \label{eq:lambda-DE-CPL}
\end{equation}
where the overall sign depends on the orientation chosen for the scalar field $\phi$. This quantity may be compared with the ekpyrotic requirement:
\begin{equation}
 \lambda_{\rm ek,req}\simeq\frac{2\Nsm}{\deff}\,.
 \label{eq:lambda-ek-req}
\end{equation}
A cyclic scalar sector model must therefore interpolate between a shallow present day slope and a very steep ekpyrotic branch. A useful slope-hierarchy diagnostic is:
\begin{equation}
 \mathcal R_\lambda\equiv\frac{\lambda_{\rm ek,req}}{\left|\lambda_{\rm DE}(a=1)\right|}\simeq\frac{2\Nsm}{\deff \left|\lambda_{\rm DE}(1)\right|}\,.
 \label{eq:slope-hierarchy}
\end{equation}
For $\Nsm=60$ and $\deff=1$, the ekpyrotic requirement is already $\lambda_{\rm ek,req}\simeq120$, whereas late-time quintessence slopes close to $\omega=-1$ are typically much smaller. This is the \textit{late-time/early-time slope-hierarchy problem for cyclic scalar models}, visualized in Fig.~\ref{fig:CPL-slope-hierarchy}. Moreover, Table~\ref{tab:CPL-distance-interpretation} summarizes the interpretation of CPL regions in the distance budget framework.

\begin{table*}[htbp]
\caption{Interpretation of CPL regions in the distance budget framework.
The table distinguishes cases where the CPL reconstruction admits a single canonical scalar-field distance from cases requiring a multifield, phantom, modified gravity, or effective fluid interpretation.}
\label{tab:CPL-distance-interpretation}
\begin{ruledtabular}
\begin{tabular}{
p{0.25\textwidth}
p{0.30\textwidth}
p{0.36\textwidth}
}
CPL behavior
&
Scalar interpretation
&
Distance-budget meaning
\\
\hline
\(w(a)>-1\) over the interval
&
canonical quintessence
&
real canonical distance \(d_{\rm DE}^{\rm CPL}\)
\\[0.4em]
\(w(a)=-1\) crossing
&
quintom, multifield, or modified gravity
&
mechanism-dependent late-time distance \(d_{\rm DE}\); no single canonical scalar description
\\[0.4em]
\(w(a)<-1\) throughout
&
phantom field or effective fluid
&
no real canonical scalar-field distance
\\[0.4em]
\(w_0>-1,\;w_a<0\)
&
possible recent crossing
&
tests whether a cyclic scalar sector can interpolate between a shallow late-time branch and a steep ekpyrotic branch
\end{tabular}
\end{ruledtabular}
\end{table*}

If the same scalar manifold controls late-time dark energy and a cyclic ekpyrotic phase, the natural generalized budget becomes
\begin{equation}
 d_{\rm cycle}=d_{\rm DE}+d_{\rm turn}+\dek+\dconv+d_b^{(\mathcal B)}+\dpost,
 \label{eq:dcycle}
\end{equation}
and it should satisfy
\begin{equation}
 d_{\rm cycle}\le d_{\rm allowed}^{(\mathcal B)}.
 \label{eq:cycle-budget}
\end{equation}
The late-time distance then reduces the distance available for the next smoothing phase:
\begin{equation}
 d_{\rm eff}^{({\rm early})}=d_{\rm allowed}^{(\mathcal B)}-d_{\rm DE}-d_{\rm turn}-\dconv-d_b^{(\mathcal B)}-\dpost\,.
 \label{eq:deff-early-with-DE}
\end{equation}
Consequently:
\begin{equation}
 \epsilon_{\rm ek}\ge\left[\frac{\Nsm}{\sqrt2\,d_{\rm eff}^{({\rm early})}}
 +\sqrt{1+\frac{\Nsm^2}{2\left[d_{\rm eff}^{({\rm early})}\right]^2}}\right]^2\,.
 \label{eq:eps-with-DE-distance}
\end{equation}
Thus, a late-time scalar motion acts as a distance debit that tightens the next ekpyrotic fast-roll limit. If the reconstructed equation of state crosses the phantom divide, the same logic applies only after specifying the appropriate multifield or modified gravity distance replacing \eqref{eq:dDE-CPL}.

\section{Limitations}
\label{sec:novelty}
The entropy-curvature relation \eqref{eq:Kfs-master} assumes approximately constant $\epsilon$ and a decoupled entropy mode during the generation stage. It should be regarded as a controlled analytic estimate, not as a replacement for full numerical perturbation evolution in a complete model. Nevertheless, it captures a robust tension: large $\epsilon$ and small field distance require either a strongly tachyonic transverse potential, strong turns, or a highly curved scalar manifold. This is the same type of geometric sensitivity encountered in multifield inflationary systems with curved target spaces \cite{Renaux-Petel:2015mga}.

Finally, the analogy with the Lyth bound should not be overstated. In inflation the field range is tied directly to $r$ through slow-roll consistency relations. In ekpyrotic and bouncing cosmology, the more robust statement is the following: large smoothing range with efficient conversion and controlled bounce implies large field-space cost, not a universal relation between $r$ and $\Delta\phi$.

\section{Conclusions}
\label{sec:conclusions}
We have formulated a phase-resolved trajectory length criterion for ekpyrotic, bouncing and cyclic cosmologies. The central quantity is the accumulated invariant path length along the actual scalar field trajectory, not in general the geodesic distance between the endpoints in field space. Since \(d_{\rm geo}\le d_{\rm traj}\), the criterion developed here is stronger than a geodesic distance requirement and should be interpreted as a sufficient, conservative criterion for small-field control, not as a model-independent necessary exclusion criterion. In particular, \(d_{\rm eff}\le0\) means that no completion is possible within the adopted finite trajectory length criterion. It should not be interpreted as a model-independent exclusion based solely on geodesic distance swampland reasoning.

The physical implications are direct. A fully small-field ekpyrotic/cyclic history tends to require ultra-fast-roll contraction, strong BKL suppression, sharp or geometrically supported entropy conversion, a short, strongly modified, or mechanism-resolved bounce, a sufficiently high EFT cutoff, and a carefully arranged entropy sector capable of producing the observed red scalar tilt. These requirements can be confronted with present and future data through $n_s$, running, non-Gaussianity, residual isocurvature, CMB tensor searches, stochastic GWs backgrounds and late-time dark energy reconstructions, and the cyclic slope hierarchy implied by a DESI-motivated CPL interpretation. In this sense the \textit{distance budget is a non-inflationary analogue of the logic behind the Lyth bound}: \textit{it does not universally predict} $r$, \textit{but it identifies the field-space cost of producing a smooth, observationally viable Universe without inflation}.

\appendix

\section{A smooth scalar-mediated toy bounce}
\label{app:toy-scalar-bounce}

\begin{figure*}[htbp]
 \centering
 \includegraphics[width=1\textwidth]{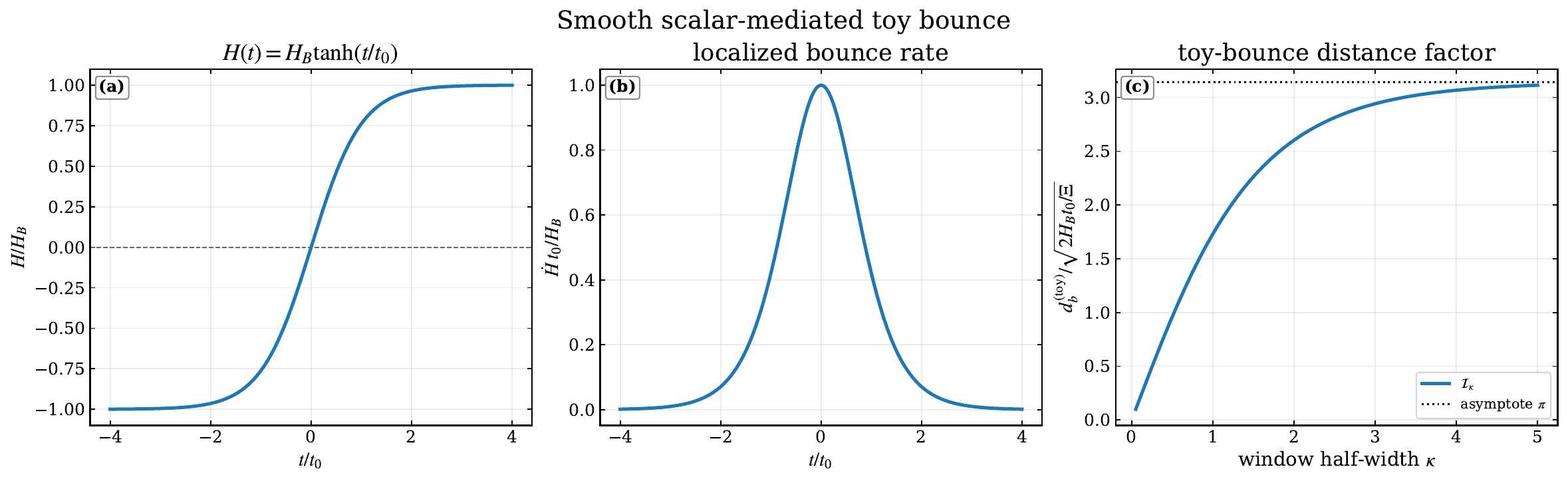}
 \caption{Toy model scalar-mediated bounce diagnostics for the ansatz \(H(t)=H_B\tanh(t/t_0)\). The panels show the normalized Hubble profile, the localized positive \(\dot H\) core, and the normalized bounce distance \(d_b^{({\rm toy})}/\sqrt{2H_Bt_0/\Xi}\) as a function of \(\kappa\), approaching the asymptotic value \(\pi\).}
 \label{fig:toy-bounce-profile}
\end{figure*}

This appendix gives a minimal worked example, summarized visually in Fig.~\ref{fig:toy-bounce-profile}, showing how the bounce distance term used in the main text is computed in an explicit scalar-mediated effective bounce. The example is not intended as a complete UV model. It is a local effective description of a smooth bounce core, similar in spirit to ghost-condensate or Galileon realizations in which the NEC-violating stage is carried by scalar dynamics \cite{Arkani-Hamed:2003pdi,Battarra:2014tga,Ijjas:2016tpn}.

Let us consider a smooth symmetric ansatz:
\begin{equation}
 H(t)=H_{B}\tanh{\left(\frac{t}{t_0}\right)}\,,
 \label{eq:toy-H}
\end{equation}
restricted to the finite core interval:
\begin{equation}
 -\kappa t_0\le t\le \kappa t_0
 \quad\land\quad
 \kappa>0\,.
\end{equation}
The bounce occurs at $t=0$, where $H=0$ and:
\begin{equation}
 \dot{H}(t)=\frac{H_B}{t_0}\sech^{2}{\left(\frac{t}{t_0}\right)}>0\,.
 \label{eq:toy-Hdot}
\end{equation}
We impose also that the effective scalar-mediated bounce relation can be written as:
\begin{equation}
 \dot H=\frac{\Xi}{2\mpl^2}\sigmadot^2
 \quad\land\quad
 \Xi>0\,,
 \label{eq:toy-Xi}
\end{equation}
with approximately constant $\Xi$ inside the bounce core. Then:
\begin{equation}
 \sigmadot(t)=\mpl\sqrt{\frac{2H_B}{\Xi t_0}}\,\sech{\left(\frac{t}{t_0}\right)}\,.
 \label{eq:toy-sigmadot}
\end{equation}
The dimensionless scalar distance consumed by the bounce is:
\begin{equation}
 d_b^{({\rm toy})}(\kappa)=\frac1{\mpl}\int_{-\kappa t_0}^{\kappa t_0}\sigmadot(t)\,dt=\sqrt{\frac{2H_Bt_0}{\Xi}}\,\mathcal I_\kappa\,,
 \label{eq:toy-db}
\end{equation}
where:
\begin{equation}
 \mathcal I_\kappa\equiv\int_{-\kappa}^{\kappa}\sech{u}\,du=4\arctan\left[\tanh\left(\frac{\kappa}{2}\right)\right]\,.
 \label{eq:toy-Ikappa}
\end{equation}
For $\kappa\to\infty$, $\mathcal I_\kappa\to\pi$, so the localized bounce contribution approaches:
\begin{equation}
 d_b^{({\rm toy})}\to\pi\sqrt{\frac{2H_Bt_0}{\Xi}}\,.
 \label{eq:toy-db-infinite}
\end{equation}
Thus, the distance cost is reduced by increasing the phenomenological effective NEC-violating coefficient $\Xi$ or by making the bounce short in the dimensionless combination $H_{B}t_{0}$.

By defining:
\begin{equation}
 I_1=\int_{-\kappa t_0}^{\kappa t_0}\sigmadot\,dt
 \quad\land\quad
 I_2=\int_{-\kappa t_0}^{\kappa t_0}\sigmadot^2\,dt\,.
\end{equation}
Using \eqref{eq:toy-sigmadot}, one gets:
\begin{equation}
 I_1=\mpl\sqrt{\frac{2H_Bt_0}{\Xi}}\,\mathcal I_\kappa
 \quad\land\quad
 I_2=\mpl^2\frac{4H_B}{\Xi}\tanh{\kappa}\,.
\end{equation}
The duration and Hubble jump over the same core interval are:
\begin{equation}
 \Delta t_b=2\kappa t_0
 \quad\land\quad
 \Delta H_b=2H_{B}\tanh{\kappa}\,.
\end{equation}
The effective coefficient computed from its definition is:
\begin{equation}
 \Xi_{\rm eff}=\frac{2\mpl^2\Delta H_b}{I_2}=\Xi\,,
\end{equation}
and the shape factor is:
\begin{equation}
 S_\kappa=\frac{I_1^2}{\Delta t_b I_2}=\frac{\mathcal I_\kappa^2}{4\kappa\tanh\kappa}\,.
 \label{eq:toy-Skappa}
\end{equation}
The general bounce distance identity is then verified explicitly:
\begin{equation}
 \left[d_b^{({\rm toy})}\right]^2=\frac{2S_\kappa\Delta H_b\Delta t_b}{\Xi_{\rm eff}}\,.
 \label{eq:toy-identity}
\end{equation}
This example also illustrates why the shape factor is useful. The function $\dot\sigma(t)$ is localized around $t=0$, and enlarging the formal interval by increasing $\kappa$ adds little field distance but increases the nominal duration. The decrease of $S_\kappa$ compensates for this bookkeeping choice.

Inserting the toy result into the mechanism-resolved distance budget gives:
\begin{equation}
 d_{\rm eff}^{({\rm toy})}=d_{\rm allowed}^{({\rm toy})}-d_{\rm conv}
 -d_{\rm post}-\sqrt{\frac{2H_Bt_0}{\Xi}}\,\mathcal I_\kappa\,.
 \label{eq:toy-deff}
\end{equation}
The smoothing criterion becomes:
\begin{equation}
 \epsilon_{\rm ek}\ge\left[\frac{\Nsm}{\sqrt2\,d_{\rm eff}^{({\rm toy})}}
 +\sqrt{1+\frac{\Nsm^2}{2[d_{\rm eff}^{({\rm toy})}]^2}}\right]^2\,,
 \label{eq:toy-epsilon-bound}
\end{equation}
provided that $d_{\rm eff}^{({\rm toy})}>0$. If the bounce is too long, too weakly modified, or too expensive in field-space distance, then $d_{\rm eff}^{({\rm toy})}$ becomes small or negative and the ekpyrotic fast-roll requirement becomes correspondingly stronger or impossible to satisfy within the assumed finite distance budget.

\begin{acknowledgments}
The author would like to thank \textit{Roksana Szwarc} for her accurate observations and comments in the initial phase of preparing the article. In addition, the author dedicates the article to \textit{Roksana Szwarc}, his parents: \textit{Bo\.zena} and \textit{Krzysztof}, his sister \textit{Dominika}, in loving memory of \textit{Grandma Stefcia}, and last but not least: \textit{Antonio}, \textit{Bob}, \textit{Bob II}, \textit{Groovy}, \textit{Milena} and \textit{Stuart}.

This article is based upon work from COST Action CA21136 – “Addressing observational tensions in cosmology with systematics and fundamental physics (\href{https://cosmoversetensions.eu/}{CosmoVerse})”, supported by COST (European Cooperation in Science and Technology).

This article is based upon work from COST Action \href{https://cosmicwispers.eu/}{COSMIC WISPers} CA21106, supported by COST (European Cooperation in Science and Technology).
\end{acknowledgments}
\section*{Data Availability}
No data were created or analyzed in this study.

\bibliographystyle{apsrev4-2}
\bibliography{main}

\end{document}